\documentclass{aa}

\usepackage{txfonts}

\usepackage{graphicx}	
\usepackage[colorlinks=true,
    linkcolor=blue,
    citecolor=blue,
    filecolor=magenta,      
    urlcolor=cyan]{hyperref}

\newcommand{\lsim}{\mathrel{\rlap{\lower 3pt \hbox{$\sim$}} \raise 2.0pt \hbox{$<$}}}
\newcommand{\gsim}{\mathrel{\rlap{\lower 3pt \hbox{$\sim$}} \raise 2.0pt \hbox{$>$}}}

\newcommand{\fr}[1]{\textcolor{black}{ #1}}
\newcommand{\frig}[1]{\textcolor{black}{ #1}}

\begin{document} 

   \title{Variability in the \fr{supermassive} black hole binary candidate SDSS J2320+0024: No evidence of periodic modulation} 
   \titlerunning{Variability in the \fr{SBHB} candidate SDSS~J2320+0024}
   \authorrunning{F. Rigamonti et al.}

   \author{Fabio Rigamonti
          \inst{1,2,}
          \inst{4} \fnmsep\thanks{fabio.rigamonti@inaf.it},
          Lorenzo Bertassi\inst{3},
          Riccardo Buscicchio\inst{3,2},
          Fabiola Cocchiararo\inst{3,2},
          Stefano Covino\inst{1,4},
          Massimo Dotti\inst{3, 1, 2},
          Alberto Sesana\inst{3, 1, 2},
          and
          Paola Severgnini\inst{1}
    }

   \institute{
            INAF - Osservatorio Astronomico di Brera, via Brera 20, I-20121 Milano, Italy
        \and
            INFN, Sezione di Milano-Bicocca, Piazza della Scienza 3, I-20126 Milano, Italy
        \and
            Universit\`a degli Studi di Milano-Bicocca, Piazza della Scienza 3, 20126 Milano, Italy
        \and
        Como Lake centre for AstroPhysics (CLAP), DiSAT, Università dell’Insubria, via Valleggio 11, 22100 Como, Italy
    }

   \date{Received XXX; accepted YYY}

\abstract 
{\fr{Supermassive} black hole binaries (\fr{SBHBs}) are a natural outcome of galaxy mergers, and they are expected to be among the loudest gravitational-wave sources at low frequencies. The source SDSS J2320+0024 was recently proposed as a promising \fr{SBHB} candidate due to a possible periodicity in its light curve and variability in the MgII emission line.
In this work, we reanalysed the optical (g \frig{r, and i} bands)  light curves of J2320+0024 within the framework of Bayesian model selection. When periodicity was searched for together with red noise, analysis of the g-band light curve reveals a peak in the posterior of the period at $\sim$290 days. The posterior profile is too broad to yield a preference for periodic models over models that include only red noise. Furthermore, the same peak is not present in the analysis of the r-band \frig{and i-band} light curve. A periodic model without red noise identified a different ($\sim$1100 days) periodicity, but this model is statistically significantly disfavoured relative to the other models tested. In summary, we find no significant evidence in favour of a true periodic signal over red-noise variability. \fr{Our analysis questions the robustness of the previously proposed periodicity and emphasises the importance of rigorous statistical treatment}. While our findings challenge the binary interpretation for J2320+0024, they do not rule it out. A statistically robust joint analysis of the photometric light curves and evolving broad-line profiles would shed further light on the true nature of this object.}


\maketitle

\section{Introduction}
\label{sec:intro}

In the current picture of hierarchical galaxy growth, \fr{supermassive} black hole (SBH) binaries (\fr{SBHB}s) are expected to form and be fairly common in our Universe as a consequence of galaxy mergers \citep[][]{BBR80,volonteri16,rg19,Derosa2019}. The study, characterisation, and identification of these systems from an electromagnetic perspective is of paramount importance in light of current, e.g., pulsar timing array \citep[PTA, ][]{PTA}, and future, e.g., Laser Interferometer Space Antenna \citep[LISA][]{AmaroSeoane2017,2022arXiv220306016A} gravitational wave (GW) missions, as these systems are expected to be among the loudest sources of GWs. Nevertheless, \fr{SBHB}s remain observationally elusive, as several challenges complicate their unambiguous identification through electromagnetic signatures. To date, despite the large number of \fr{SBHB} candidates that have been put forward, no definite observational confirmation of any has yet been provided. 
The identification of \fr{SBHB}s relies on indirect signatures, either through searches for distinctive spectral features \citep[][]{tsalmantza,eracleous,Ju13,Shen13,Wang17} or through photometric variability in their light curve \citep[][]{valtonen,Ackermann15,Graham15,Li2016,Charisi16,Sandrinelli16,Sandrinelli18,Severgnini18,Li+2019,LiuGez+2019,Chen+2020,Serafinelli+20,Covino2019,Luo2025}.  

The first approach is associated with \fr{SBHB}s that have separations larger than $0.01$ pc. At such distances, typically larger than the Roche lobe of the system, individual SBHs still retain their own broad-line regions (BLRs) \citep[][]{montuori11,montuori12}. 
In these cases, we expect broad emission lines (BELs) to be shifted in frequency with respect to their narrow emission lines (NELs), evolving in time over a binary orbital period. 
However, asymmetric emission line profiles can also be explained by the complex dynamics and morphology of the BLR \citep[][]{eracleous97,Jovanovic_2010,Storchi-Bergmann,Rigamonti2025,Sottocorno2025}. 
The presence of an\fr{SBHB} can be tested by measuring the expected drift in wavelengths over a binary orbital period, i.e. by observing the candidate for $\approx 10 {\rm yr}$ or longer \footnote{A faster test, requiring an observational campaign on timescales comparable to typical reverberation mapping observation, was proposed by \cite{Gaskell88} and more recently has been quantified by \cite{Dotti2023}.}. 

The second approach is more suitable for binaries at closer separations, where the individual BLRs are either truncated or shared by \fr{both black holes bound in the binary system}. 
In such configurations, periodic variability on orbital timescales is expected, possibly due to periodic feeding from the circumbinary disc \citep[][]{HMH08,Tiede2024}, Doppler-boosted emission \citep{DHS15}, or periodic gravitational lensing \citep[][]{doraziolense18, davelaar22a}. However, even in such situations, convincing evidence that these sources are indeed \fr{SBHB}s is still missing, due both to theoretical predictions for their emission and to the plausibility of alternative interpretations, such as precession in jets and in the inner part of the accretion discs \citep[][]{Sandrinelli16,Britzen2018}. Solving this ambiguity requires alternative approaches, encompassing time-domain and spectral analyses, \citep[e.g.][]{Bertassi2025} to robustly test the binary hypothesis. 
Finally, the most challenging complication in the identification of genuine \fr{SBHB}s from light-curve analysis arises from intrinsic quasar variability. 
Indeed, the light curves of accreting active galactic nuclei (AGNs) generally show correlated red-noise variability, with power spectra well described by a $\nu^{-2}$ power law that flattens at low frequency, i.e. a damped random walk \citep[DRW;][]{Kelly2009}. Such variability can mimic quasi-periodicities, making genuine periodicities difficult to identify unambiguously\citep[][]{vaughan, liu,Covino2019}.
Multiple objects that initially appeared to show periodicity were later revealed to be non-periodic \citep[][]{Graham15,liu,Jiang2022,Dotti2023}. Established approaches to assess source periodicity now rely on Bayesian inference \citep[][]{Covino2020,Zhu2020,Hubner_2022}.

The source SDSS J2320+0024 (hereafter J2320+0024) has been proposed to be an \fr{SBHB} candidate due to a detected peak (T$\simeq 273$ days) in its Lomb-Scargle periodogram \citep[][]{Fatovic2023}, possibly coupled with variability in its Mg II broad emission line \citep[][]{Fatovic2025}. However, to date, no rigorous assessment of the significance of the observed periodicity has been provided. 
In particular, a comparison with quasar red noise and an evaluation within a robust Bayesian framework are still lacking. In this study, our objective was to test the binary nature of J2320+0024 by focusing on the periodicity detected in its light curve. 

In Sec.~\ref{sec:data_sample}, we briefly present the archival datasets of the \fr{SBHB} candidate analysed in this paper. We then describe our approach to analysing the light curve of J2320+0024 in Sec.~\ref{sec:modelling}, ranking different noise and periodic models against each other. 
Finally, in Sec.~\ref{sec:discussion}, we discuss our results and present our conclusions.

\section{Data}
\label{sec:data_sample}

J2320 + 0024 is a redshift system $z=1.05$, identified through a long-period variability search in the Sloan Digital Sky Survey Stripe 82 Standards Catalog \citep[SDSS S82,][]{York00,Jiang2014,Ivezic2007,Thanjavur2021}. It was proposed as an \fr{SBHB} candidate by \cite{Fatovic2023} based on time-resolved analysis of its light curve and, and was later followed up with spectral observations\citep[][]{Fatovic2025}. To assess the robustness of the periodicity detection in the light curve of J2320+0024, we collected g-, r-, and \frig{i-}band magnitudes from the SDSS, Pan-STARRS1 \citep[PS1,][]{Chambers2016}, and Zwicky Transient Facility \citep[ZTF,][]{Graham2019_ZTF,Bellm2019_ZTF} surveys. Following \cite{Fatovic2023}, we calibrated the data from the different surveys with the median SDSS flux to empirically account for instrumental offsets. In the following, we present the analysis on the g-band, while the corresponding results for the r- \frig{and i-band} are reported in Appendix~\ref{App:analysis_on_r_band} and Appendix~\ref{App:analysis_on_i_band} respectively. 

\begin{table*}
	\centering

 \caption{Summary of  model parameters for the g-band light-curve fit of SDSS J2320+0024. }
	\label{tab:parameters}
	{\begin{tabular}{llcccccc} 
        \hline\vspace{-0.75em}\\
		Description & Name & Prior range & DRW & Modified DRW & QPO & Periodic \vspace{0.3em}\\
        \hline\vspace{-1em}\\
        \multicolumn{7}{c}{\tiny \bf 50th, 16th, 84th Percentiles}\\ \hline\vspace{-0.5em}\\
		Scale length & $\log_{10}{\tau}$ & $[-3,1.7]$ & $-1.3^{+0.3}_{-0.2}$ & $-1.6^{+0.3}_{-0.2}$ & $-1.3^{+0.4}_{-0.3}$ & -\vspace{0.2em}\\
		Dispersion & $\log_{10}{\sigma}$ & $[-3,1.7]$ & $0.8^{+0.1}_{-0.1}$ & $0.9^{+0.1}_{-0.1}$ & $0.8^{+0.1}_{-0.1}$ & $0.1^{+0.3}_{-0.2}$ \vspace{0.2em}\\
        Slope & $\gamma$ & $[0.5,5]$ & - & $1.4^{+0.4}_{-0.4}$ & - & - \vspace{0.2em}\\
		Period & $\log_{10}{T}$ & $[-2,2]$ & - & - & $0.2^{+1.3}_{-1.3}$ & $-0.406^{+0.002}_{-0.002}$\vspace{0.3em}\\
        \hline\vspace{-1em}\\
        \multicolumn{7}{c}{\tiny \bf Maximum Likelihood}\\ \hline\vspace{-0.5em}\\
		Scale length & $\log_{10}{\tau}$ & $[-3,1.7]$ & $-1.35$ & $-1.65$ & $-0.74$\vspace{0.2em}\\
		Dispersion & $\log_{10}{\sigma}$ & $[-3,1.7]$ & $0.79$ & $0.94$ & $0.51$ & $0.02$ \vspace{0.2em}\\
        Slope & $\gamma$ & $[0.5,5]$ & - & $1.51$ & - & - \vspace{0.2em}\\
		Period & $\log_{10}{T}$ & $[-2,2]$ & - & - & $-0.99$ & $-0.41$ \vspace{0.2em}\\
        
        \hline\vspace{-0.75em}\\
            log evidence & $\log{Z}$ & $-$ & -139.15 & -139.0 & -138.90 & -238.60 &\vspace{0.2em}\\
        \hline
        
	\end{tabular}
    \tablefoot{From left to right, the columns provide a brief description of each parameter, its reference name as used in this work, the assumed prior range, and the best-fit values with their credibility intervals for all the models. The first table block reports parameters estimated as the median, 16th, and 84th percentiles of the posterior distribution, while the second table block reports the parameter set that maximises the likelihood. All priors are log-uniform and all the parameters are in dimensionless units since the data were standardised before fitting. The last row reports the Bayesian log evidence of each model, clearly disfavouring the Periodic model.}}

\end{table*}

\section{Modelling of the light curve}
\label{sec:modelling}
As discussed in Sec.~\ref{sec:intro}, quasi-stellar objects (QSOs) are variable sources whose intrinsic variability is typically well described by a DRW \citep[][]{Kelly2009} model. This model can mimic a spurious periodic signal \citep[][]{vaughan}, as a random realisation of a DRW with a correlation timescale $\tau$ can produce few periodic cycles spanning a timescale $\gsim \tau$. Therefore, it is extremely challenging to confirm periodicities that are not much shorter than the data span, as a robust claim typically requires several ($O(10)$) periods. Consequently, robust detection of periodic variability, whether associated with the presence of an \fr{SBHB} \citep[][]{liu} or with other repeating processes \citep[][]{Covino2019}, is difficult and requires a cautious approach. We analysed the light curves of J2320+0024 within a fully Bayesian framework, modelling the data as realisations of a Gaussian process (GP) and performing model selection among multiple models.

This approach has been proposed previously in the literature 
\citep[][]{Covino2020,Zhu2020,Covino2022,Hubner_2022} and constitutes the most reliable method for quasar light-curve characterisation.

\subsection{Gaussian processes and Bayesian inference}
\label{Sec:gp}

Bayesian analysis of light curves typically makes use of the GP formalism \citep[][]{Rasmussen_2006,Roberts2012,Angus2018}. 
Gaussian processes describe distributions over functions and, in their discretised version, correspond to multivariate Gaussian distributions. 
This framework offers a flexible approach for modelling unknown time series through non-parametric models capable of capturing a large family of functions with two parameters: a kernel and a mean function.
The kernel choice, which defines the GP covariance matrix, regulates the overall shapes and smoothness of samples, i.e. functions. Kernels can have different functional forms, usually depending on a few free (hyper)-parameters, allowing the modelling of either periodic or non-periodic light curves. In Bayesian inference, identifying the kernel parameters that best describe the observed data requires evaluating the GP log marginal likelihood, which is given by \cite{Rasmussen_2006}

\begin{equation}
\label{eq:likelihood}
\log{\mathcal{L}}(\mathbf{D}|\boldsymbol{\Omega}) = 
 -\frac{1}{2} (\mathbf{D} - \boldsymbol{\mu})^T 
\widetilde{{\boldsymbol{\Sigma}}}^{-1} (\mathbf{D} - \boldsymbol{\mu}) - \frac{1}{2}\log{|\widetilde{\boldsymbol{\Sigma}}|}  - \frac{N}{2}\log{2\pi}.
\end{equation}

Here, $\mathbf{D}$ denotes the observed magnitude, $\boldsymbol{\mu}$ is the mean function of the GP, $N$ is the number of observations, $\widetilde{\boldsymbol{\Sigma}}$ is the covariance matrix computed as the squared sum of the GP kernel $\boldsymbol{\Sigma}$ and the diagonal term given by the error measurements, while $|\widetilde{\boldsymbol{\Sigma}}|$ is the covariance matrix determinant. To minimise the number of free parameters in the model, we standardised $\mathbf{D}$  by removing the mean and dividing by the standard deviation\footnote{The mean g-band magnitude is 21.616 while the standard deviation is 0.319.}) allowing us to fix $\boldsymbol{\mu}=0$.

We compared four different kernels typically used to describe periodic and non-periodic light curves: the exponential kernel DRW, the modified DRW kernel, the QPO kernel \fr{, i.e. a superposition of noise and a periodic signal}, and the periodic kernel. We refer the reader to \cite{Zhu2020} for a detailed discussion. 

The DRW model assumes an exponential kernel described by 
\begin{equation}
    \label{eq:DRW}
    \Sigma_{ij} = \frac{1}{2} \sigma^2 \tau \exp{\left(-\frac{|t_i-t_j|}{\tau}\right)},
\end{equation}
where $\sigma$ is the intrinsic variance, $\tau$ is the damping timescale, and $t_i$ is the time at which the $i-$th observation was taken. As with the magnitudes, the observational times were also standardised\footnote{The mean modified Julian date (MJD) is 56766.93, while the standard deviation is 2834.764.}. This reduced numerical issues when inverting the covariance matrix $\boldsymbol{\Sigma}$. 

The DRW model can be generalised by including an additional free parameter $\gamma>0$ as an exponent to the $|t_i-t_j|/\tau$ term. This modification accounts for quasar red noise that deviates from the DRW model. This model, referred to as the modified DRW, is characterised by the covariance matrix   

\begin{equation}
    \label{eq:modified_DRW}
    \Sigma_{ij} = \frac{1}{2} \sigma^2 \tau \exp{\left[-\left(\frac{|t_i-t_j|}{\tau}\right)^{\gamma}\right]}.
\end{equation}
Eqs.~\eqref{eq:DRW},\eqref{eq:modified_DRW} represent pure noise models. When accounting for periodicity, kernels typically include terms proportional to an oscillating function. We define the quasi-periodic oscillation (QPO) kernel as

\begin{equation}
    \label{eq:QPO}
    \Sigma_{ij} = \frac{1}{2} \sigma^2 \tau \exp{\left(-\frac{|t_i-t_j|}{\tau}\right)} \cos{\left(\frac{2\pi|t_i-t_j|}{T}\right)},
\end{equation}
which contains a multiplicative periodic modulation with period $T$ to the DRW model. The QPO model accounts for a superposition of noise and a periodic signal and reduces to the DRW model in the limit $T\to\infty$. Gaussian processes are a flexible, intrinsically stochastic approach to model functions. The shape of the covariance matrix regulates the smoothness of the GP realisations but is not reflected in a unique shape of the output function (i.e. it is not deterministic). For example, the covariance in Eq.~\ref{eq:QPO} does not restrict the model to purely cosine-like periodicities, but allows for more general signals with probabilistic periodic correlations. In the absence of noise, the QPO kernel describes a purely oscillatory process:

\begin{equation}
    \label{eq:cosine}
    \Sigma_{ij} = \frac{1}{2} \sigma^2 \cos{\left(\frac{2\pi|t_i-t_j|}{T}\right)},
\end{equation}
which we refer to as the periodic model. 
The final goal of this work was to measure the robustness of the presence of a periodic signal in the light curve of J2320+0024. Bayesian statistics is well suited for this task, as it allows precise model selection through the computation of the model evidence. The Bayesian evidence of a model ($Z$) is the normalisation factor of the posterior distribution and enables the selection of the preferred model given the same dataset. Given two competing models, $H_i$ and $H_j$, the Bayes ratio is
\begin{equation}
    B_{ij} = \frac{p(D|H_i)}{p(D|H_j)} = \frac{Z_i}{Z_j},
\end{equation}
where no a priori preference for $H_i$ over $H_j$  and $Z_i/Z_j$ is assumed. The ratio of the model evidences obtained by further marginalising over the entire parameter spaces for $H_i$ and $H_j$ is called the Bayes factor.  
If $B_{ij}$ is > 1 (< 1), then model $H_i$ is favoured (disfavoured) compared to $H_j$. We refer to \cite{Jeffreys_probability} for a detailed classification and to Cocchiararo et al. (in prep.) for a more detailed discussion of GP inference in our astrophysical context. 

\subsection{Fit and model selection}
Since our goal was to estimate the Bayesian evidence of the proposed models, we adopted a nested sampling algorithm for parameter inference \citep[][]{Skilling2004}. In particular, following previous works \citep[e.g.][]{Rigamonti2022,Rigamonti2023}, we employed the \textsc{python}-based implementation \textsc{raynest} \citep[][]{Del_Pozzo_CPNest}, which efficiently computes the marginal likelihood (Eq.~\ref{eq:likelihood}) using the \textsc{GPyTorch} package \citep[see][]{GpyTorch}\footnote{While Eq.~\eqref{eq:likelihood} is a marginal-likelihood for a single GP, it is instead to be interpreted as a likelihood when exploring GPs with different parameters.}. 
We performed a parameter estimation with 1000 live points, assuming log-uniform priors for all parameters, except $\gamma$, which we assumed to have a uniform prior. The prior ranges adopted for all models are reported in  Tab.~\ref{tab:parameters}. 

\begin{figure}
    \centering
    \includegraphics[width=\linewidth,trim={0 0 0 1.25cm}]{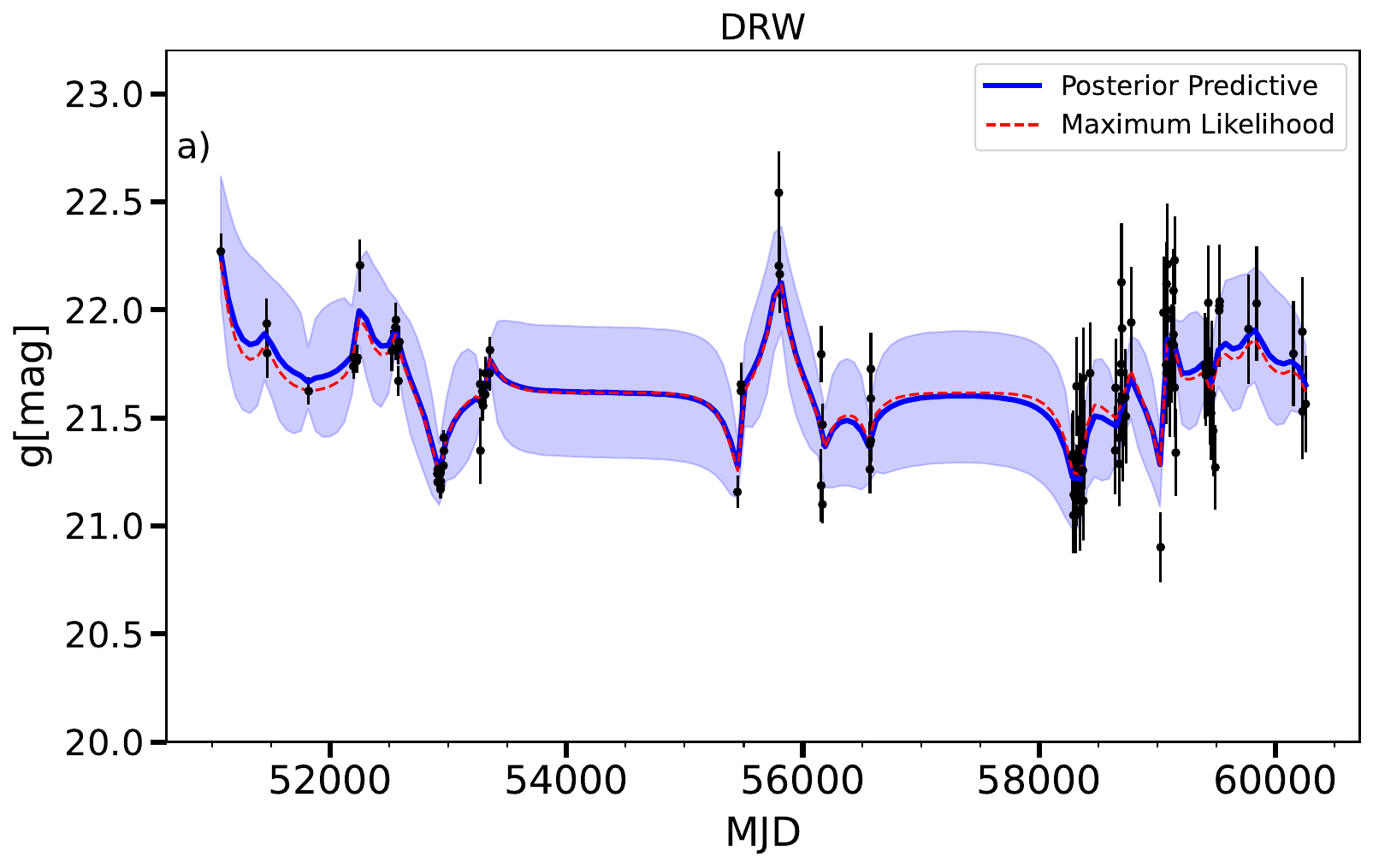}
    \includegraphics[width=\linewidth]{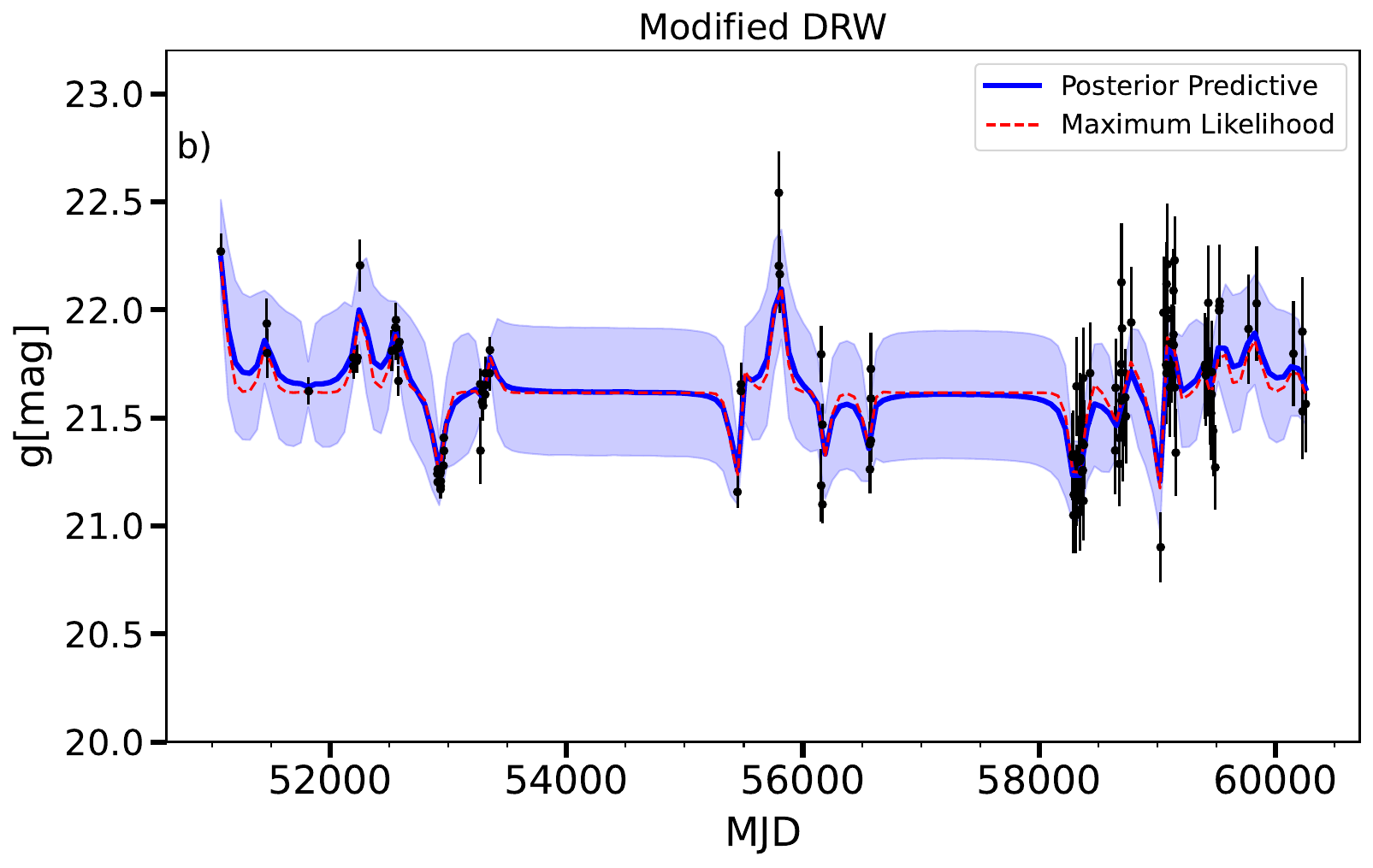}
    \includegraphics[width=\linewidth]{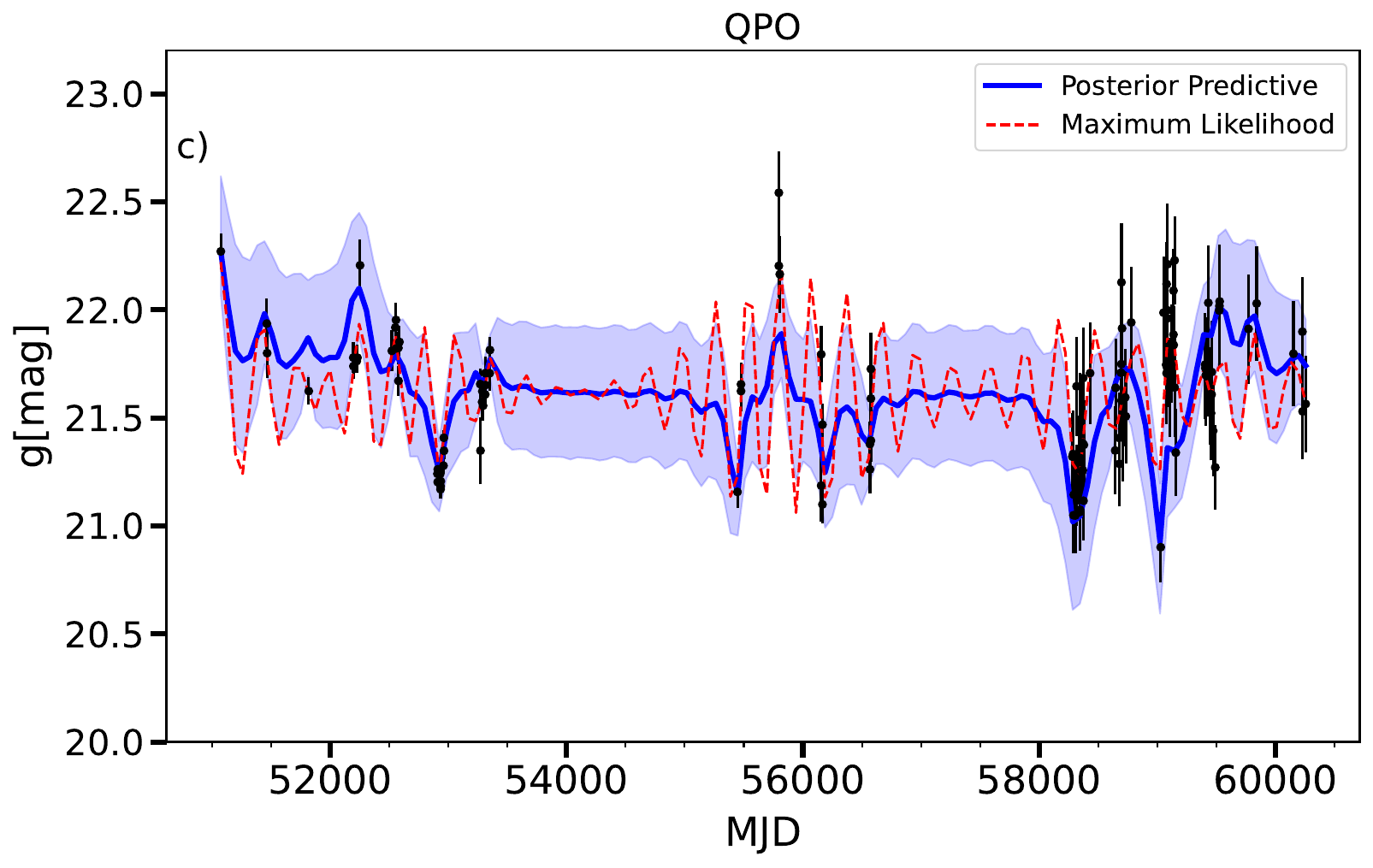}
    \includegraphics[width=\linewidth,trim={0 0.8cm 0 0}]{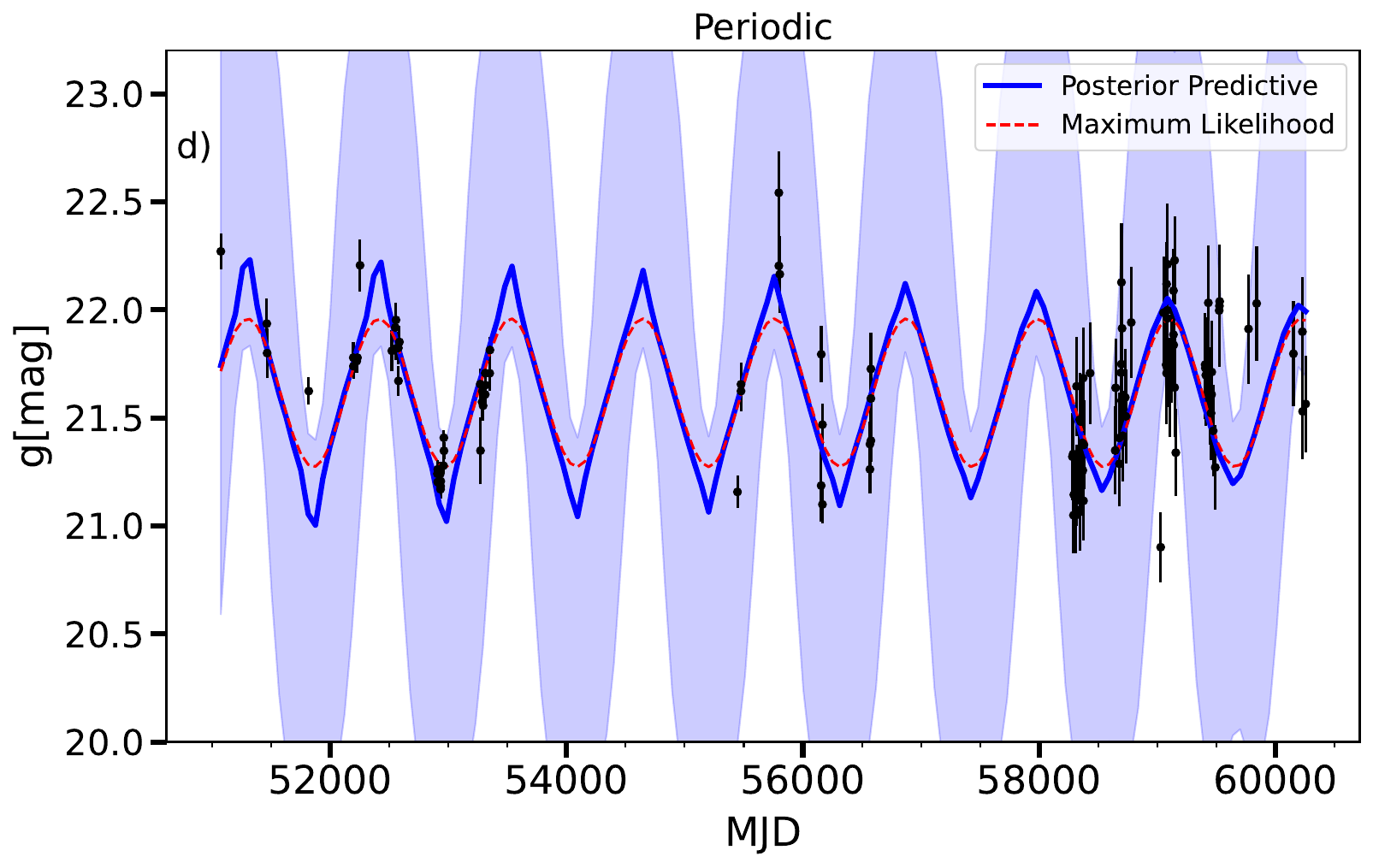}
    \caption{Best-fit to the g-band light curve of J2320+0024. Black points represent the data with their errors, the solid blue  line and shaded area show the median, 16th, and 84th percentiles of the posterior distributions, and the red dashed line indicates the maximum-likelihood model. From top to bottom: DRW (a), modified DRW (b), QPO (c), and Periodic (d) . See Table~\ref{tab:parameters} for the estimated best-fit parameters.}
    \label{fig:best_fit}
\end{figure}

\begin{figure}
    \centering
    \includegraphics[width=\linewidth]{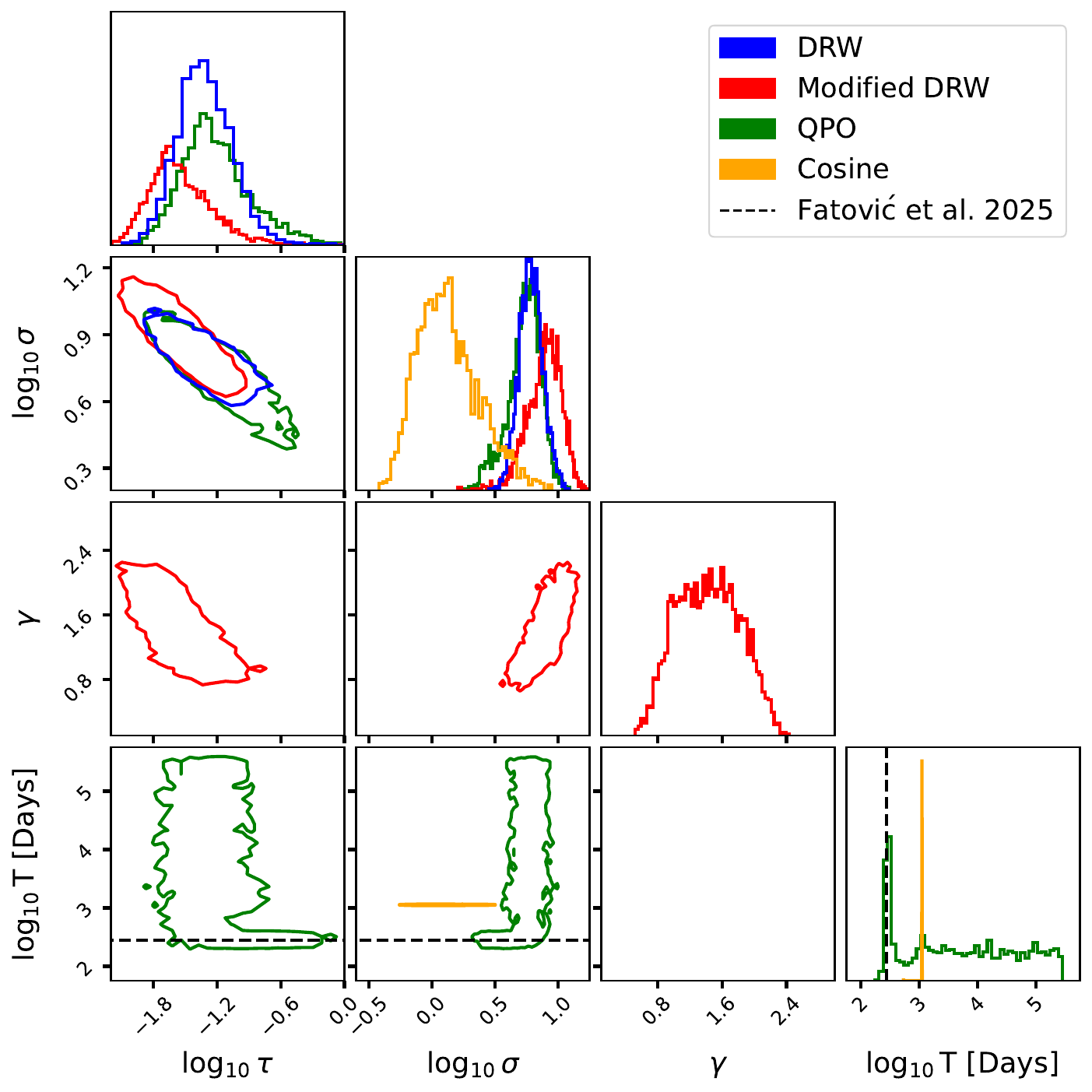}
    \caption{Corner plot of the posteriors for the different models. The DRW model is shown in blue, modified DRW in red, QPO in green, and Periodic in orange. Contours are drawn to include 90\% of the posterior probability. The dashed black lines indicate the period reported in \cite{Fatovic2025}. For clarity, although the fit is performed on standardised data, the marginal posterior distribution of the period is shown in units of $\log_{10}$ days.}
    \label{fig:Corner_plot}
\end{figure}

Fig.~\ref{fig:best_fit} shows the best-fit light curves with associated errors for the four models discussed in Sec.~\ref{Sec:gp}: DRW (a), modified DRW (b), QPO (c), and periodic (d). In all plots, we report the data with their associated uncertainties; the posterior predictive median, 16th, and 84th percentiles; and the maximum likelihood model. \frig{More precisely, the posterior predictive median and percentiles were computed by sampling 100 GP realisations for a fixed set of posterior-sampled kernel hyperparameters. This allowed effective marginalisation over the kernel hyperparameters, preserving and propagating all information in their posterior distribution}.
The best-fit model parameters (either percentiles or likelihood maxima), together with their corresponding Bayesian evidence values, are reported in Tab.~\ref{tab:parameters}. It is important to clarify that, while the maximum likelihood models (i.e. dashed red lines) are obtained using the "maximum likelihood" parameters reported in Tab.~\ref{tab:parameters}, this is not the case for the median, 16th, and 84th percentile models. In agreement with our Bayesian approach, the blue lines and the shaded areas presented in Fig.~\ref{fig:best_fit} represent the median, 16th, and 84th percentiles computed by evaluating the GP model on all parameter combinations describing the posterior distribution. \fr{Therefore, the blue shaded area reported in Fig.~\ref{fig:best_fit} encapsulates and propagates three different uncertainties: data variance (i.e. measured errors), kernel variance (i.e. the kernel value at $t_i=t_j$), and the posterior distribution of the kernel parameters.} \frig{The combination and propagation of these three sources of error in the posterior predictive are highly non-linear and difficult to disentangle. For instance, measurement errors contribute to the production of a smoother model, less dependent on outliers and highly uncertain points, while the kernel variance and the hyperparameter uncertainty simultaneously contribute to the specific shape of the posterior predictive. These effects depend on the chosen model. For example, the periodic model exhibits a larger uncertainty in the posterior predictive with respect to the others, due to a combination of factors. Specifically, this model, unlike the others, fits the data poorly and therefore requires a larger $\sigma$. The large $\sigma$ is needed due to the specific shape of the covariance matrix (see Eq.~\ref{eq:cosine}), which suppresses kernel variance in the regions where the cosine is close to zero.}Fig.~\ref{fig:Corner_plot} displays a comparison of the marginalised posterior distribution of the DRW (blue), modified DRW (red), QPO (green), and periodic (orange). The contours of the 2D probability densities were chosen to include 90\% of the probability.

Combining the results shown in Fig.~\ref{fig:best_fit}, Fig.~\ref{fig:Corner_plot}, and Tab.~\ref{tab:parameters}, allows us to highlight some key insights. Visually, little difference is observed between the DRW (Fig.~\ref{fig:best_fit}-a) and the modified DRW (Fig.~\ref{fig:best_fit}-b) models. The Bayes factor between the two models is close to 1, indicating that there is no strong evidence in favour of either model. We stress that the DRW model corresponds to the modified DRW model with $\gamma=1$. In principle, instead of considering $\gamma$ as a free parameter, we could have fixed it to a value different from 1 (e.g. $\gamma\simeq1.5$). Such a model would produce only slightly larger evidence (i.e. $\log{Z}\simeq138.5$) and would still not indicate a strong preference for either model. Nevertheless, the comparison is useful: the estimated scale length and dispersion are shifted to smaller and larger values, respectively. This indicates that estimates of the $\tau$ and $\sigma$ parameters depend on the high-frequency slope of the assumed power spectrum, possibly suggesting that $\gamma$ might be included to provide better estimates of the physical parameters \citep[e.g. black hole mass and accretion rate,][]{Arevalo2023,Arevalo2024} that regulate the light curve. 

The QPO model exhibits an interesting behaviour. As shown in Fig.~\ref{fig:Corner_plot}, its posterior distribution peaks at values of $\tau$ and $\sigma$ highly compatible with those obtained from the DRW model. However, the period $T$ not only shows a peaked distribution corresponding to $\log_{10}{T}\simeq-1.0$ (i.e. $T\simeq 290$ days), but also features a pronounced tail extending to larger values. This suggests that although the peak at $\log_{10}{T}\simeq-1.0$ is well defined and sharp, the majority of the cumulative distribution lies at longer periods. Consequently, for values of $T$ comparable or longer than the observational timespan, the QPO model becomes insensitive to the oscillatory terms. From a Bayesian statistics perspective, this is not problematic: when a strong periodic signal is present, models with ineffective oscillatory terms contribute negligibly to the posterior and do not influence the final best-fit model estimate. However, this is not the case for J2320+0024. Indeed, as shown in Fig.~\ref{fig:best_fit}-c, the shape of the QPO posterior distribution is also responsible for the difference between the maximum likelihood and median models.
 
The red curve demonstrates fast oscillations (i.e. $\log_{10}{T}\simeq-1\simeq 290$ days), whereas the blue line is much closer to a DRW model. Moreover, the Bayes ratio reveals no significant preference between the (modified) DRW and the QPO models, precluding strong claims in favour of the latter. These findings suggest a lack of robust evidence for a periodic signal in the light curve of J2320-0024.

To compare our results with those found in \cite{Fatovic2025} and to assess the possible impact of overfitting on the QPO model, we also fit a periodic model. We note that, unlike the other models considered in this work, the evidence for the periodic model is significantly lower. This clearly suggests that there is no statistical support for preferring the periodic model over the alternatives given the current data.
Thus, there is no strong evidence that the light curve of J2320+0024 can be described by a purely periodic signal, highlighting the importance of red-noise intrinsic variability. Interestingly, the marginalised posterior distribution reveals extremely sharp period peaks ($T\simeq 1100$ days) with relative errors smaller than a few percent. 
Notably, the only peak in the period of the periodic model does not match the main peak identified by the QPO model; rather, it overlaps with a secondary, smaller peak. Nested sampling algorithms used with the periodic model detect other peaks, including one that corresponds to the periodicity reported in \citet[][]{Fatovic2025}. However, these smaller peaks have significantly lower likelihoods and are rejected by our approach. Since \citet[][]{Fatovic2025} adopt a maximum likelihood approach with the period search initialised near $T\simeq273$ days, the reported period may correspond to a local rather than a global likelihood maximum. Similar conclusions arise from analysing the histogram of the residuals (see Appendix~\ref{app:histograms}) across the four models. Although the Kolmogorov-Smirnov test \citep[][]{Smirnov} indicates that the residuals for the DRW and modified DRW models always resemble a Gaussian distribution (i.e., p-value $> 0.05$), this is not the case for the periodic model and is only partially true for the QPO model (i.e. only when using the best-fit model).

\section{Discussion and conclusion}
\label{sec:discussion}
J2320+0024 has been proposed as a candidate hosting an \fr{SBHB} based on modulation of its light curve \citep[][]{Fatovic2023}, \fr{who report a period of 278 days based on a Lomb-Scargle periodogram analysis of the g, r, and \frig{i} bands, exceeding the false alarm probability threshold.} However, our results, obtained via a rigorous Bayesian approach, find no evidence for a periodic signal in the data. This highlights the importance of applying statistically robust methods to assess the nature of variability in a given source \fr{and the exposes the limitations of periodogram analyses for detecting true periodicity in the presence of red noise}. Nevertheless, the absence of a periodic signal in the light curve of J2320+0024 does not necessarily rule out the presence of an \fr{SBHB} in this source. Our result suggests that even if such a periodic signal exists, the current dataset lacks sufficient coverage to detect it robustly.

We also find period peaks for the QPO and periodic models that correspond to $T=290$ days and $T=1100$ days, respectively, when translated into physical units. \cite{Fatovic2023,Fatovic2025} report a periodicity of $T\simeq270$ days, much closer to the value of our QPO model. We note that a period of $1100$ days is almost an integer multiple of $T=270$ days (i.e. close to four times larger). Interestingly, when our analysis is extended to the r-band \citep[i.e. the same reported in][]{Fatovic2025}, the short timescale periodicity ($T=290$ days) disappears, while the longer, statistically unfavoured, peak at $T\simeq1100$ days persists. \frig{Different behaviours are also observed in the i-band, where the short timescale periodicity of the QPO is not detected, and a double peak appears in the periodic model.} These variations in period estimation between adjacent broad bands further support the absence of a robust signal in the light curve of J2320+0024.  
  
We consider J2320+0024 a very interesting source deserving further follow-up observations and analysis, particularly to better characterise the spectral variability reported in \citet[][]{Fatovic2025}. This behaviour was explained within the framework of the Popovi\'c, Simi\'c, Kova\v{c}evi\'c, Ili\'c (PoSKI) model model \citep[][]{Popovic2021}. The model assumes the presence of an \fr{SBHB} surrounded by a common \fr{circumbinary} BLR, illuminated by emission from the two mini-discs around each binary component, \fr{with only the less massive black hole retaining its own BLR}. In this scenario, the variable blue or red wing observed in MgII is attributed to binary orbital motion, \fr{particularly to  Doppler-boosted emission from the less massive black hole's BLR}, and is expected to reflect the periodicity observed in the light curve. While intriguing, our results challenge this scenario. Indeed, \fr{as} \cite{Fatovic2025} note, alternative explanations for the observed MgII variability exist. As discussed in Sec.~\ref{sec:intro}, physical mechanisms, such as non-symmetric BLR \citep[][]{Storchi-Bergmann,Pancoast2014,Raimundo2020,Rigamonti2025}, can produce distorted broad emission profiles, possibly explaining the features observed in the MgII line of J2320+0024. However, explaining the observed variation over time remains more challenging. The most significant variation occurs between the SDSS spectrum, obtained approximately years ago, and the more recent Magellan or Gemini spectra. Over such long timescales, comparable to BLR dynamical timescales, variations in BEL shape might be expected, possibly due to changes in the morphology or dynamics of the overall BLR. Furthermore, the more recent Magellan and Gemini spectra show exhibit variations among themselves, comparable to variations from a non-symmetric BLR reverberating after DRW-like variability of the AGN \citep[][]{Sottocorno2025}. 

We stress that further photometric and spectroscopic follow-up observations are necessary to clarify the nature of J2320+0024. In particular, as mentioned in Sec.~\ref{sec:intro}, simultaneous modelling of repeated photometric and spectroscopic observations could reveal the true nature of this source. For example, the light curve modelling presented here could be extended to include a consistent response from a perturbed BLR \citep[e.g.][]{Rigamonti2025,Sottocorno2025} to provide a detailed comparison with models or tests based on the \fr{SBHB} scenario \citep[][]{Popovic2021,Bertassi2025}.

\begin{acknowledgements}
We thank the anonymous referee for their comments and suggestions that helped us to improve the quality of the paper.

FR acknowledges the support from the Next Generation EU funds within the National Recovery and Resilience Plan (PNRR), Mission 4 - Education and Research, Component 2 - From Research to Business (M4C2), Investment Line 3.1 - Strengthening and creation of Research Infrastructures, Project IR0000012 – “CTA+ - Cherenkov Telescope Array Plus.

MD acknowledge funding from MIUR under the grant
PRIN 2017-MB8AEZ, financial support from ICSC – Centro Nazionale di Ricerca in High Performance Computing, Big Data and Quantum Computing, funded by European Union – NextGenerationEU, and support by the Italian Ministry for Research and University (MUR) under Grant 'Progetto Dipartimenti di Eccellenza 2023-2027' (BiCoQ).
We acknowledge a financial contribution from the Bando Ricerca Fondamentale INAF 2022 Large Grant, {\textit{Dual and binary supermassive black holes in the multi-messenger era: from galaxy mergers to gravitational waves.}}
RB acknowledges support from the ICSC National Research Center funded by NextGenerationEU, and the Italian Space Agency grant Phase A activity for LISA mission, Agreement n.2017-29-H.0.
The data underlying this article will be shared on reasonable request to the corresponding author. AS acknowledges the financial support provided under the European Union’s H2020 ERC CoG "B Massive" (Grant Agreement: 818691) and AdG "PINGU" (Grant Agreement: 101142079)
\end{acknowledgements}

\bibliographystyle{aa} 
\bibliography{main}


\begin{appendix}

\section{Analysis on the r-band}
\label{App:analysis_on_r_band}
\begin{figure}
    \centering
    \includegraphics[width=\linewidth,trim={0 0 0 1.25cm}]{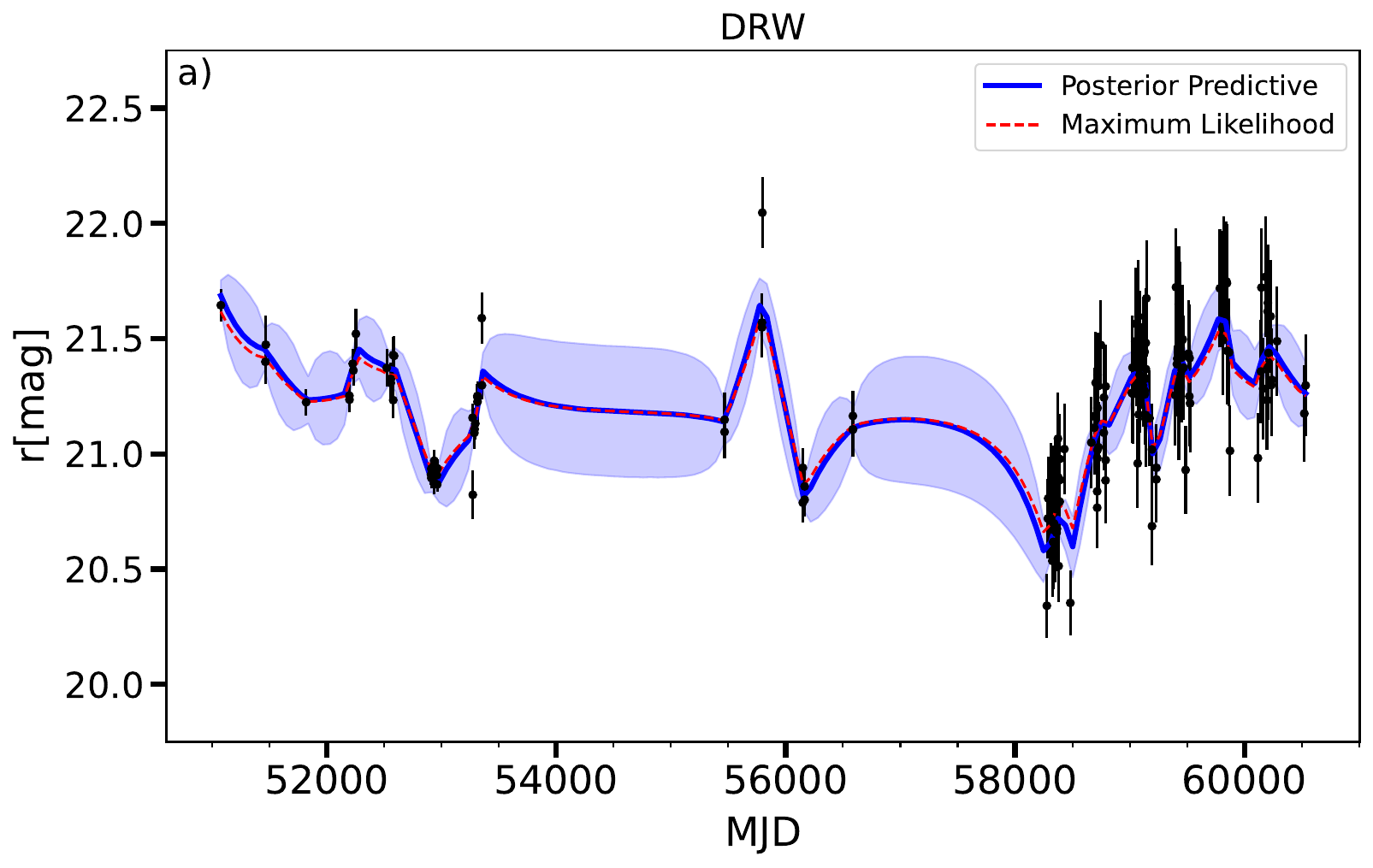}
    \includegraphics[width=\linewidth]{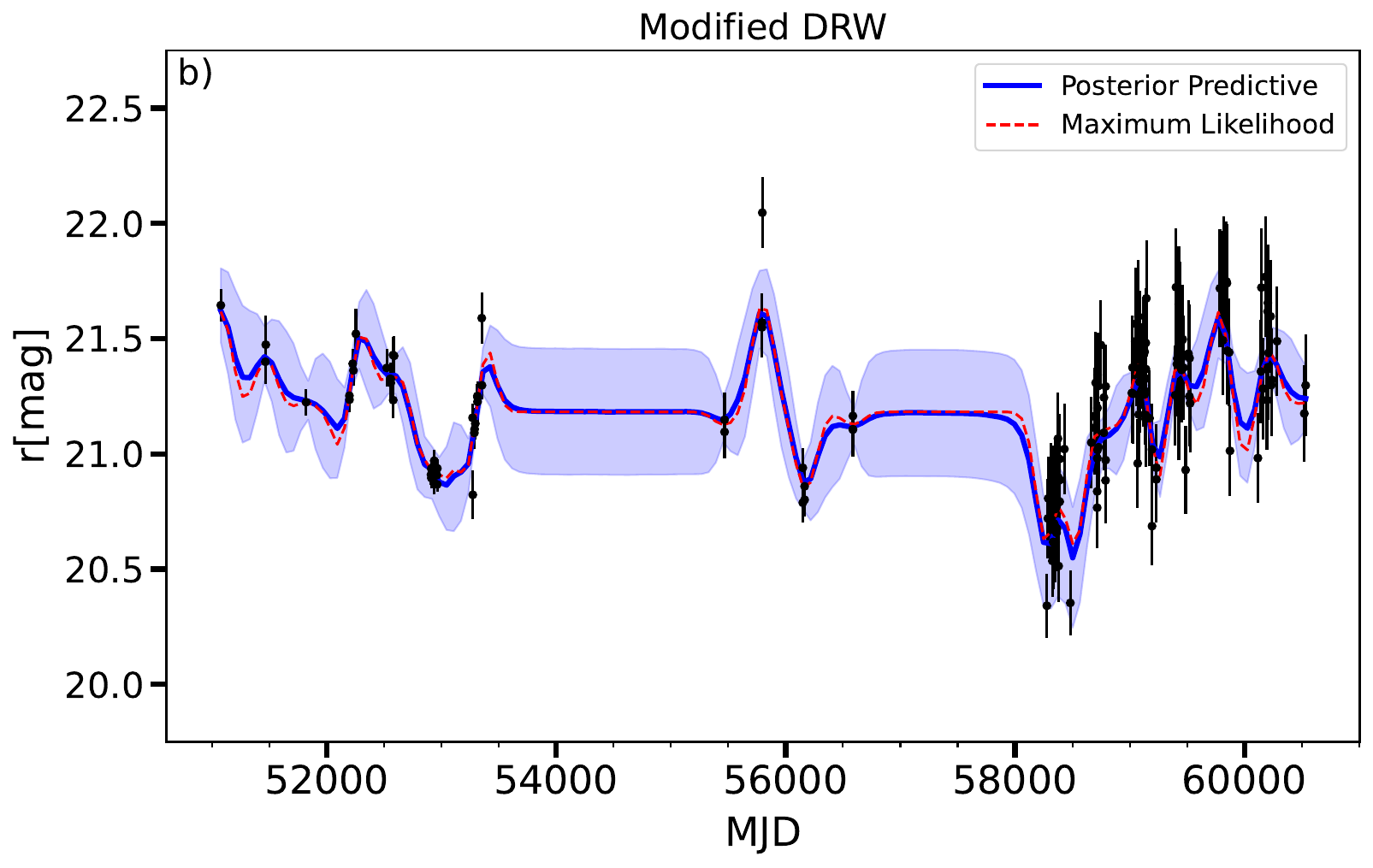}
    \includegraphics[width=\linewidth]{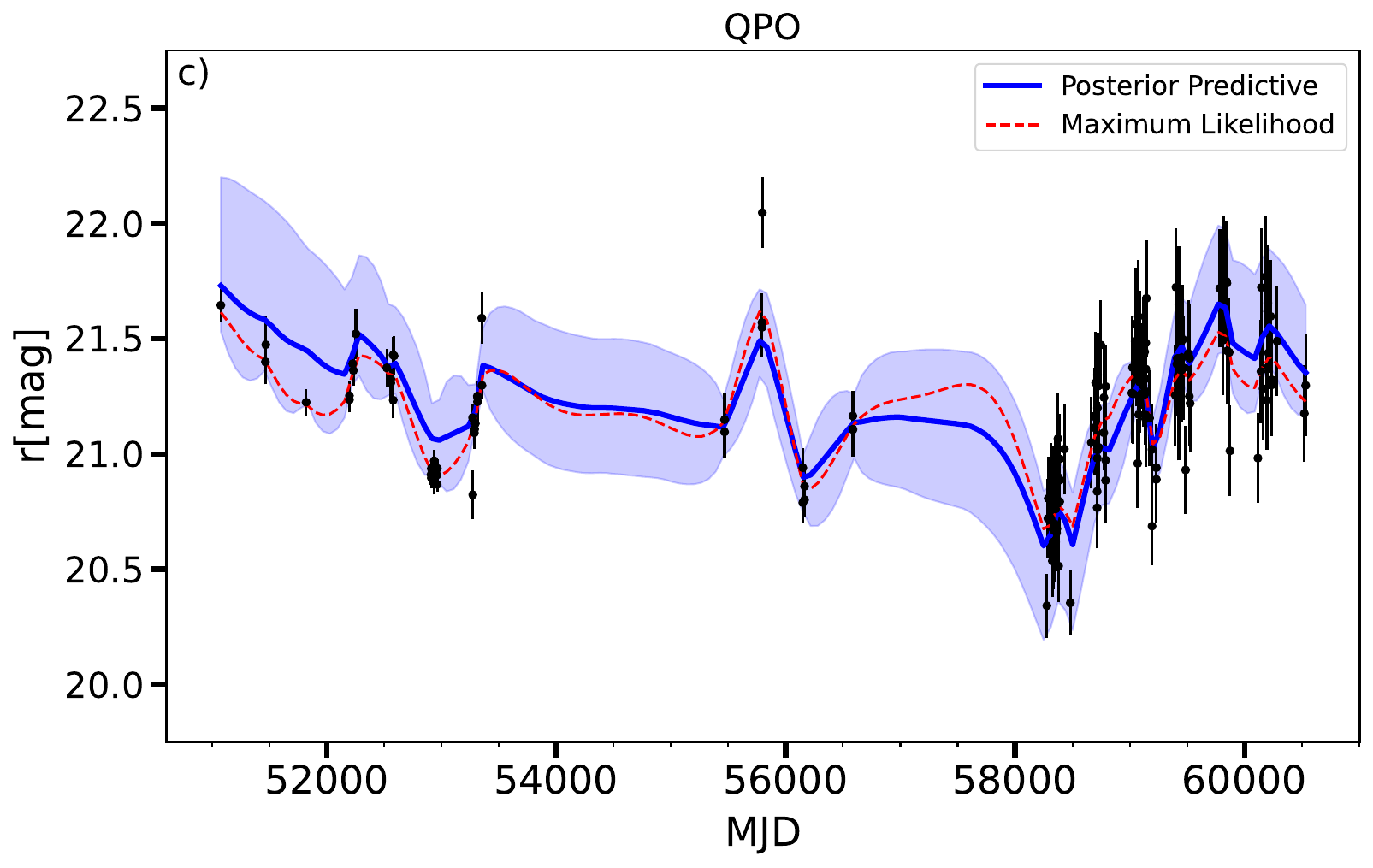}
    \includegraphics[width=\linewidth,trim={0 0.8cm 0 0}]{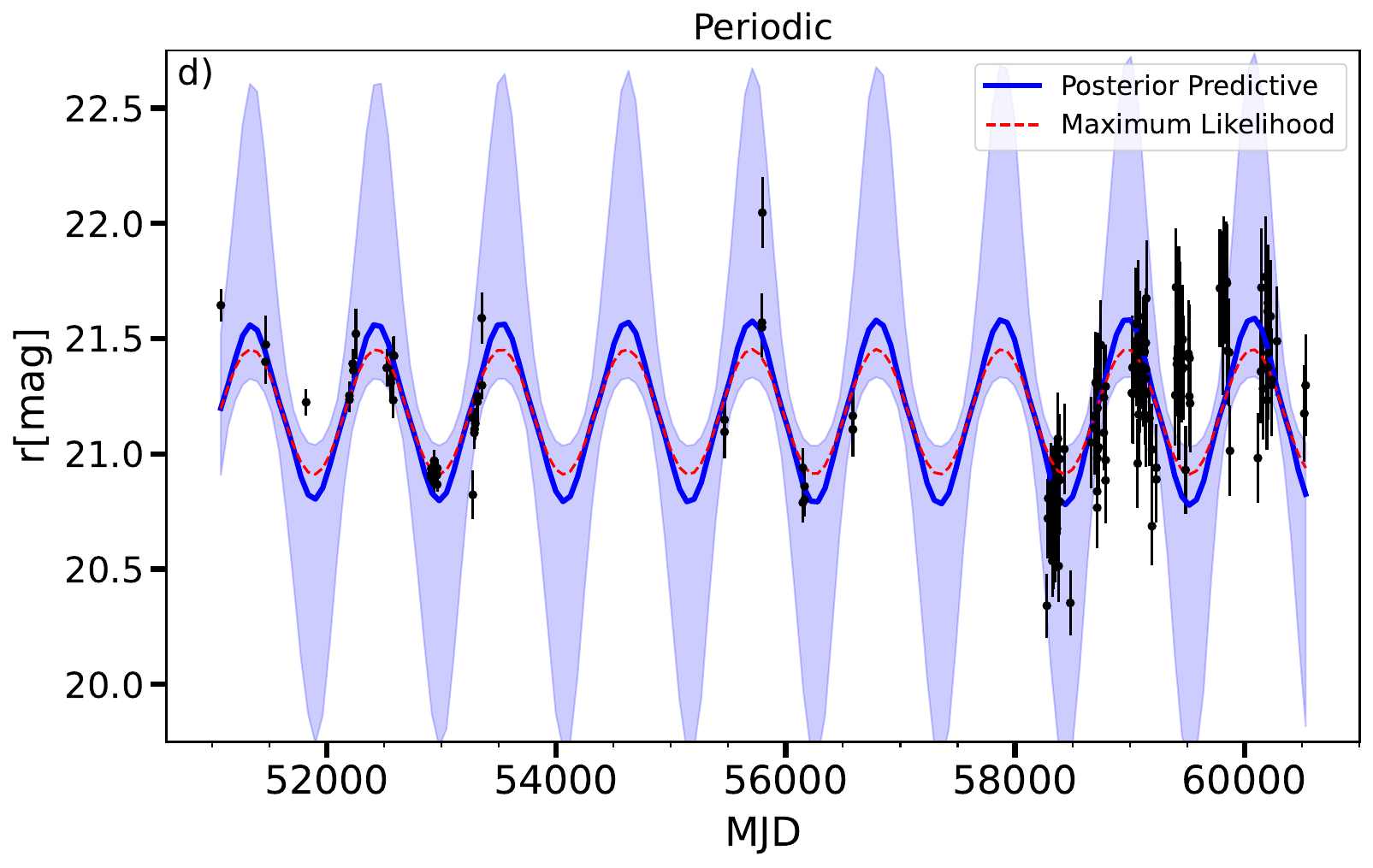}
    \caption{Best-fit for the r-band light curve of J2320+0024. The black points represent the data with their errors, the blue solid line and the shaded area represent the median, 16th, and 84th percentiles from the posterior distributions, and the red dashed line refers to the maximum likelihood model. From top to bottom: DRW (a), modified DRW (b), QPO (c), and Periodic (d), we refer to Tab.~\ref{tab:parameters} for the estimated best-fit parameters.}
    \label{fig:best_fit_rband}
\end{figure}

In this section, we repeat the same analysis of Sec.~\ref{sec:modelling} on the r-band data of J2320+0024. Also in this case, we tested four different models: DRW, Modified DRW, QPO, and Periodic reporting their best-fit light curves and parameters in Fig.~\ref{fig:best_fit_rband} and Tab.~\ref{tab:parameters_rband} respectively\footnote{Also in this case, we standardised the data using a mean r-band magnitude of 21.182 with a standard deviation of 0.319 and
a mean modified Julian date (MJD) of 57602.992 with a standard deviation of 2650.236.}.

\begin{figure}
    \centering
    \includegraphics[width=\linewidth]{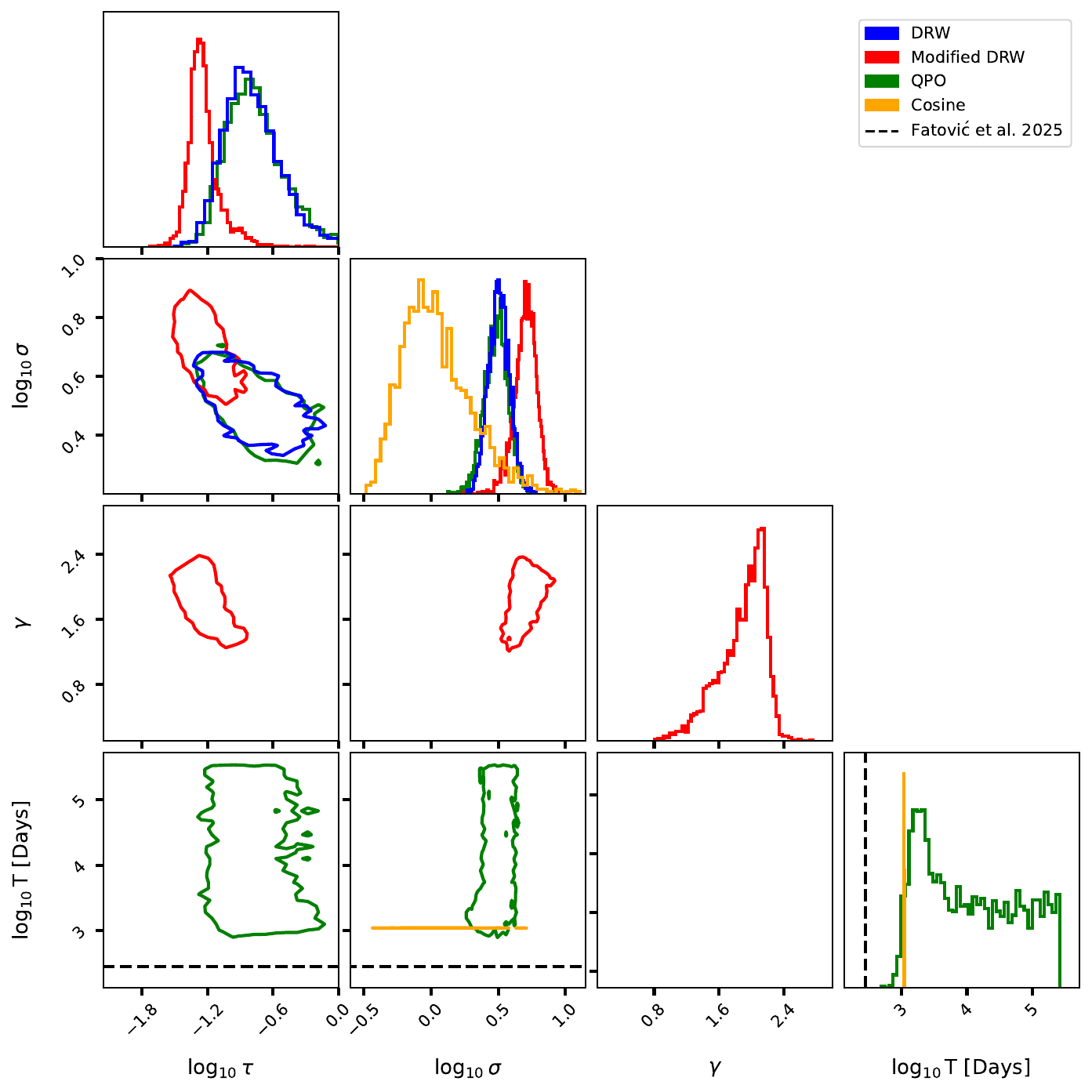}
    \caption{Corner plot of the posteriors of the different models. DRW is blue, modified DRW is red, QPO is green, and Periodic is orange. Contours are drawn at a level to include 90\% of the posterior probability, while the black dashed lines refer to the period reported in \cite[][]{Fatovic2023}. For reasons of clarity, although the fit is performed on standardised data, we report the marginal posterior distribution of the period in units of $\log_{10}$ Days. Please note the different range of $\log_{10}{T}$ in the corner plot. No short ($\simeq273$ days) oscillations are identified by the nested sampling algorithm. }
    \label{fig:Corner_plot_rband}
\end{figure}

The overall considerations from the analysis are similar to those discussed for the g-band. The Periodic model is strongly unfavoured, while the DRW, Modified DRW, and QPO have comparable evidence. Notably, in this case, the Bayes ratios between these three models point toward a slight preference for the Modified DRW. Interestingly, while the period found in the Periodic model remains close to that found in the g-band, the main peak in the period of the QPO model is larger and no longer consistent with the one reported by \cite[][]{Fatovic2025} and found by us in the g-band (see Figs.~\ref{fig:Corner_plot},~\ref{fig:Corner_plot_rband}). Such a large variation in the measured periodicities among adjacent bands further probes the lack of a robust periodicity detection in J2320+0024

\begin{table*}
	\centering

 \caption{Summary of the different model parameters for the fit to the r-band light curve of SDSS J2320+0024. }
	\label{tab:parameters_rband}
	{\begin{tabular}{llcccccc} 
        \hline\vspace{-0.75em}\\
		Description & Name & Prior range & DRW & Modified DRW & QPO & Periodic \vspace{0.3em}\\
        \hline\vspace{-1em}\\
        \multicolumn{7}{c}{\tiny \bf 50th, 16th, 84th Percentiles}\\ \hline\vspace{-0.5em}\\
		Scale length & $\log_{10}{\tau}$ & $[-3,1.7]$ & $-0.8^{+0.3}_{-0.2}$ & $-1.3^{+0.1}_{-0.1}$ & $-0.8^{+0.4}_{-0.3}$ & -\vspace{0.2em}\\
		Dispersion & $\log_{10}{\sigma}$ & $[-3,1.7]$ & $0.5^{+0.1}_{-0.1}$ & $0.7^{+0.1}_{-0.1}$ & $0.5^{+0.1}_{-0.1}$ & $0.01^{+0.3}_{-0.2}$ \vspace{0.2em}\\
		Slope & $\gamma$ & $[0.5,5]$ & - & $2.0^{+0.2}_{-0.4}$ & - & - \vspace{0.2em}\\
		Period & $\log_{10}{T}$ & $[-2,2]$ & - & - & $0.5^{+1}_{-0.7}$ & $-0.395^{+0.001}_{-0.001}$\vspace{0.3em}\\
        \hline\vspace{-1em}\\
        \multicolumn{7}{c}{\tiny \bf Maximum Likelihood}\\ \hline\vspace{-0.5em}\\
		Scale length & $\log_{10}{\tau}$ & $[-3,1.7]$ & $-0.90$ & $-1.30$ & $-0.73$ & -\vspace{0.2em}\\
		Dispersion & $\log_{10}{\sigma}$ & $[-3,1.7]$ & $0.51$ & $0.74$ & $0.44$ & $-0.06$ \vspace{0.2em}\\
		Slope & $\gamma$ & $[0.5,5]$ & - & $2.1$ & - & - \vspace{0.2em}\\
		Period & $\log_{10}{T}$ & $[-2,2]$ & - & - & $-0.20$ & $-0.39$ \vspace{0.2em}\\
        
        \hline\vspace{-0.75em}\\
            log evidence & $\log{Z}$ & $-$ & -163.0 & -161.4 & -163.4 & -275.1 &\vspace{0.2em}\\
        \hline
        
	\end{tabular}
    \tablefoot{From left to right, the columns provide a brief description of each parameter, its reference name as used in this work, the assumed prior range, and the best-fit values with their credibility intervals for all the different models. The first table block reports the parameters estimated as the median, 16th, and 84th percentiles of the posterior distribution, while the second table block reports the set of parameters that maximise the likelihood. All the priors are log-uniform, and all the parameters are in dimensionless units since the data have been standardised before fitting. The last row reports the Bayesian log evidence of each model clearly disfavoring the Periodic model.}}

\end{table*}

\section{\frig{Analysis on the i-band}}
\label{App:analysis_on_i_band}

\begin{figure}
    \centering
    \includegraphics[width=\linewidth,trim={0 0 0 1.25cm}]{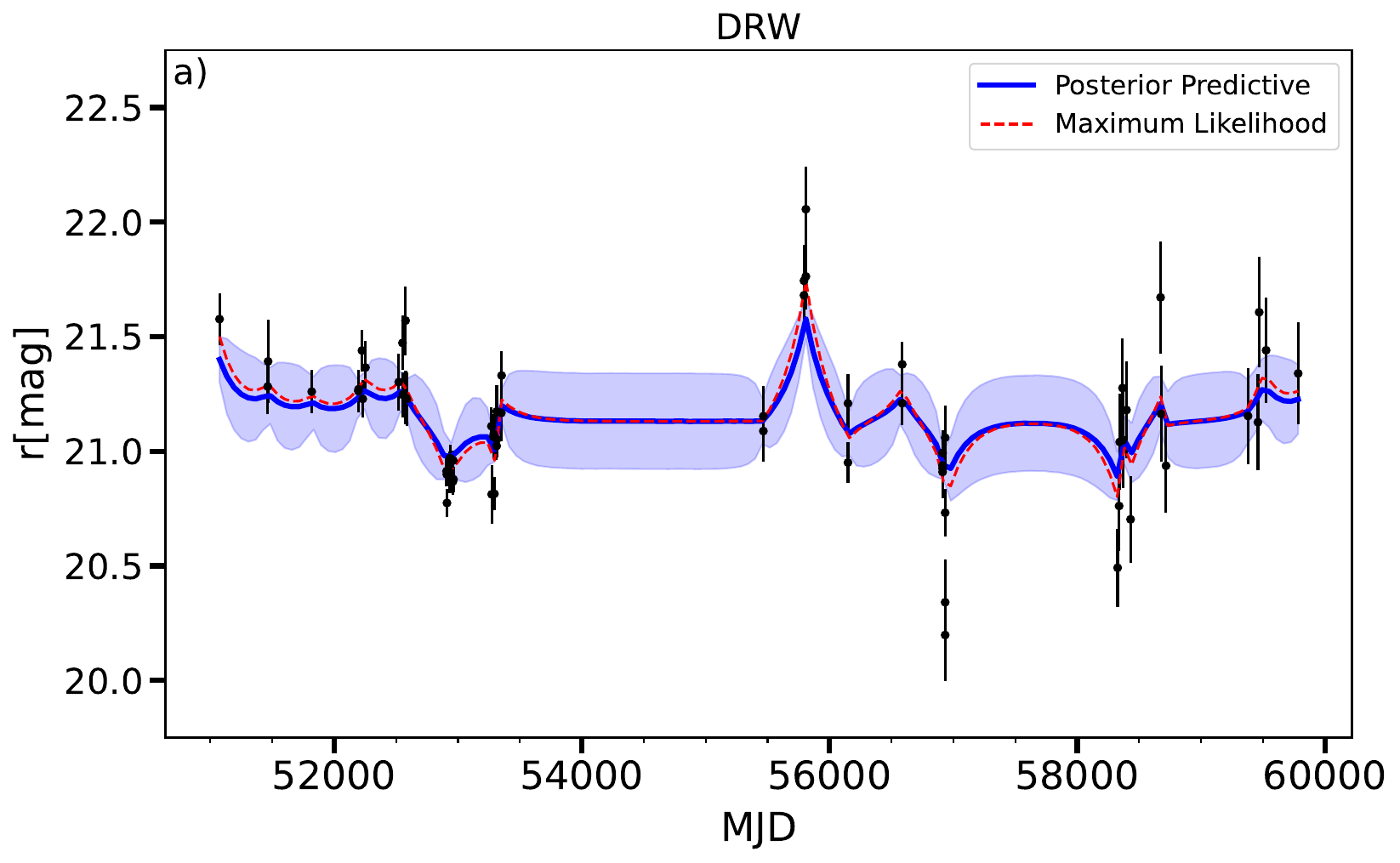}
    \includegraphics[width=\linewidth]{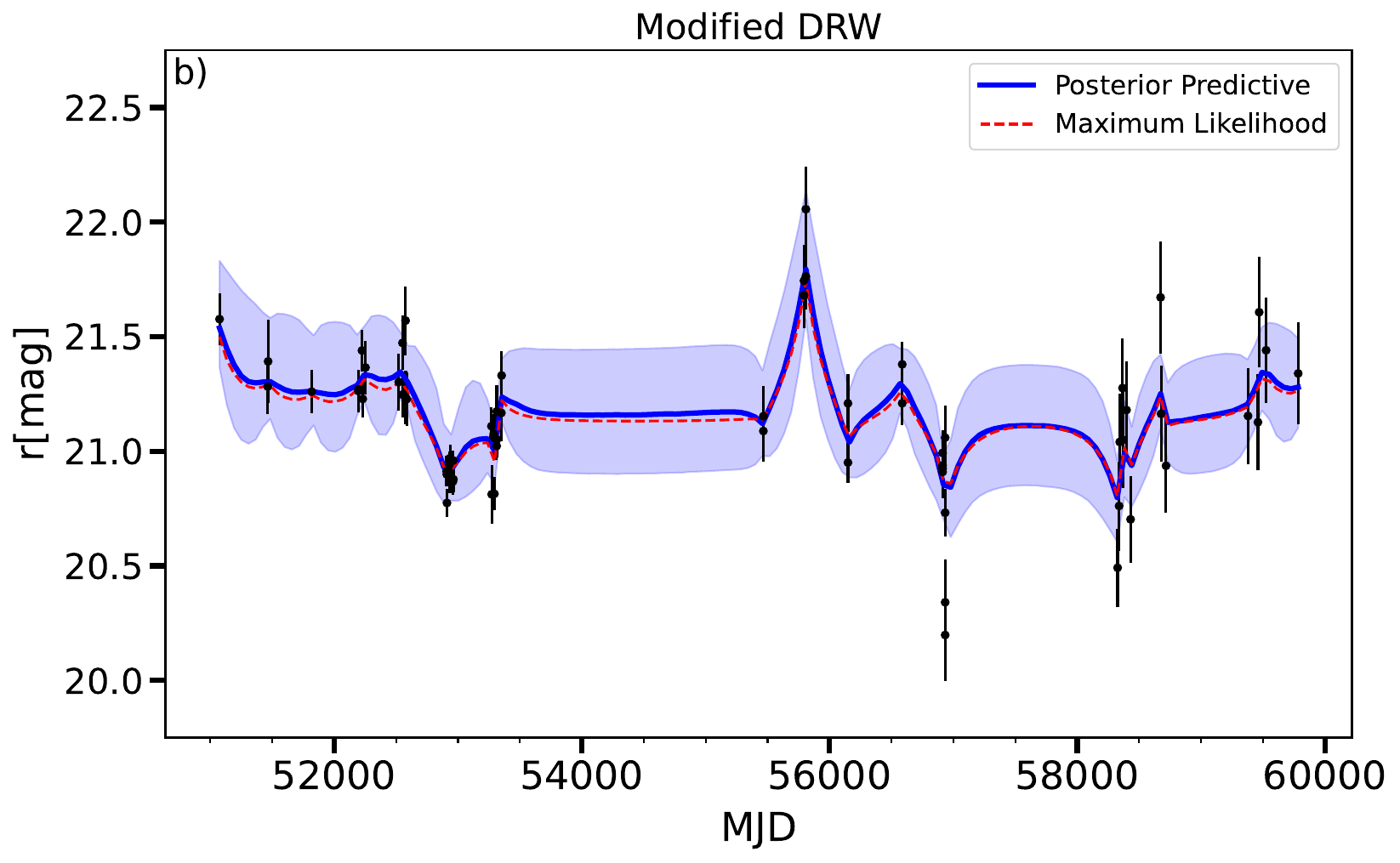}
    \includegraphics[width=\linewidth]{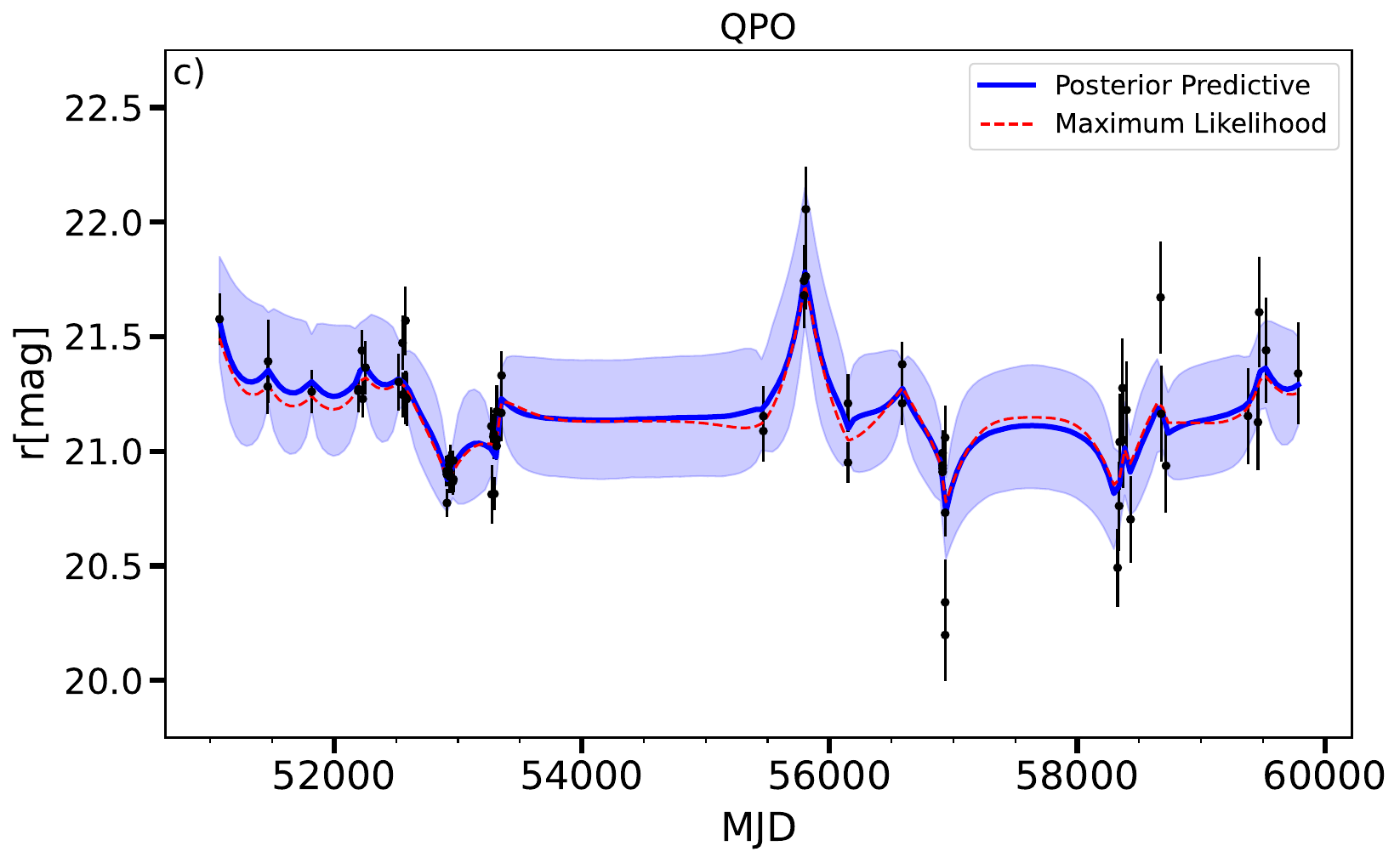}
    \includegraphics[width=\linewidth,trim={0 0.8cm 0 0}]{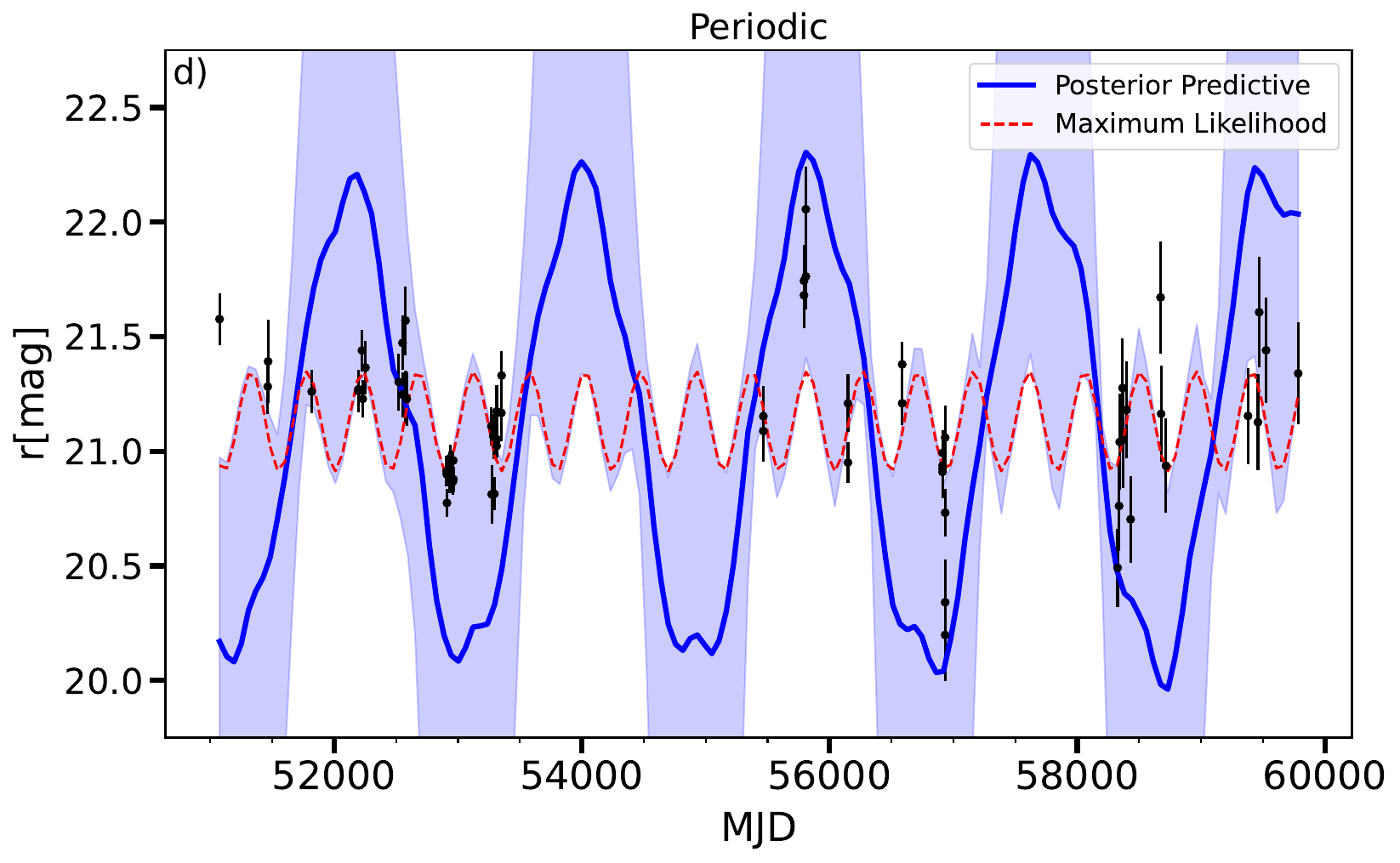}
    \caption{Best-fit for the i-band light curve of J2320+0024. The black points represent the data with their errors, the blue solid line and the shaded area represent the median, 16th, and 84th percentiles from the posterior distributions, and the red dashed line refers to the maximum likelihood model. From top to bottom: DRW (a), modified DRW (b), QPO (c), and Periodic (d), we refer to Tab.~\ref{tab:parameters} for the estimated best-fit parameters.}
    \label{fig:best_fit_iband}
\end{figure}

\frig{Here we repeat the same analysis performed in Sec.~\ref{sec:modelling} and Appendix~\ref{App:analysis_on_r_band} focusing on i-band data. The best-fit light curves for the four models are plotted with their uncertainties in Fig.~\ref{fig:best_fit_iband}, while the parameters describing the models are reported in Tab.~\ref{tab:parameters_iband}. 
\footnote{\frig{In this case, we rescale the data using a mean i-band magnitude of 21.131 (standard deviation of 0.318) and a mean MJD of 54963.152 (standard deviation of 2588.802).}}}

\begin{figure}
    \centering
    \includegraphics[width=\linewidth]{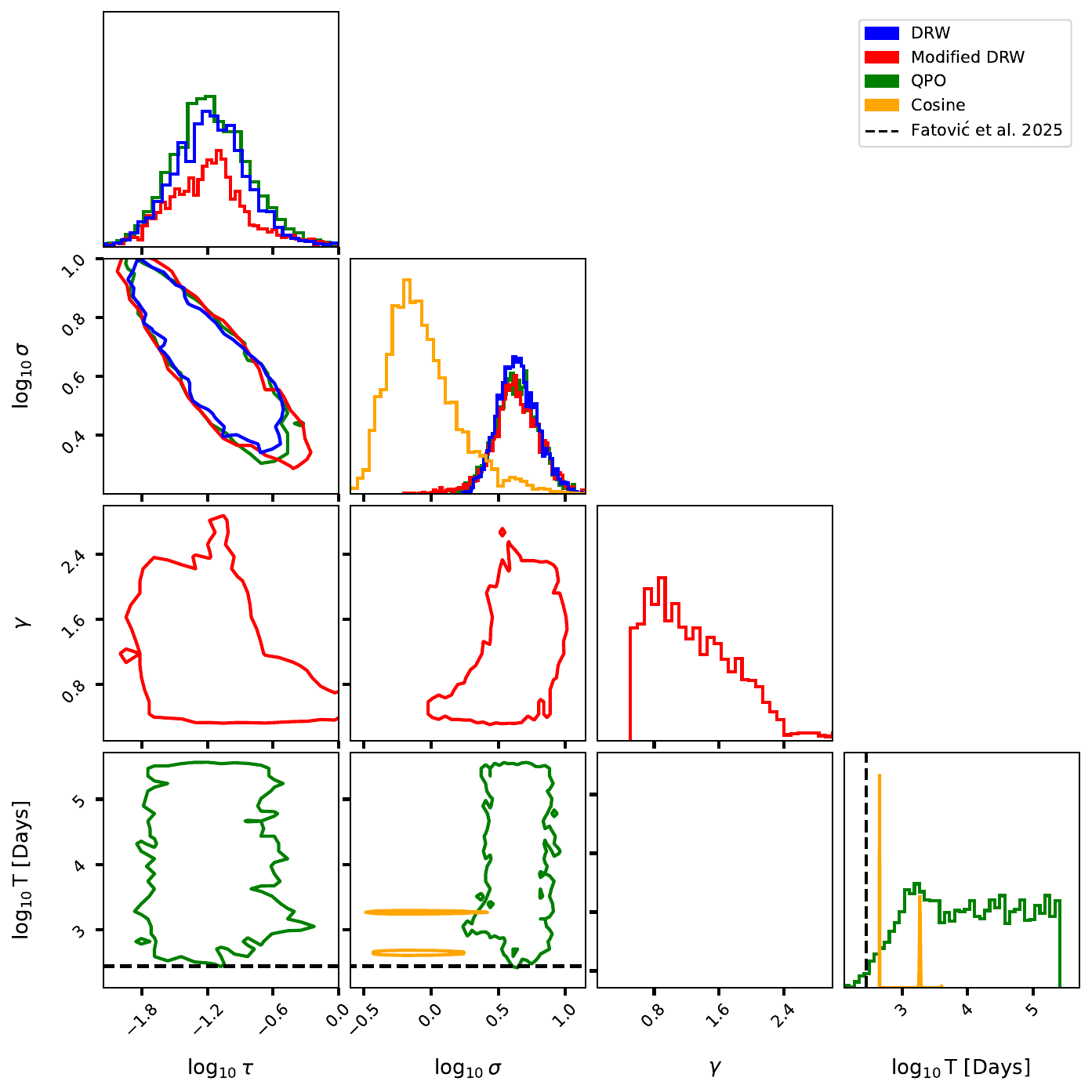}
    \caption{Corner plot of the posteriors of the different models. DRW is blue, modified DRW is red, QPO is green, and Periodic is orange. Contours are drawn at a level to include 90\% of the posterior probability, while the black dashed lines refer to the period reported in \cite[][]{Fatovic2023}. For reasons of clarity, although the fit is performed on standardised data, we report the marginal posterior distribution of the period in units of $\log_{10}$ Days. Please note the different range of $\log_{10}{T}$ in the corner plot. No short ($\simeq273$ days) oscillations are identified by the nested sampling algorithm. }
    \label{fig:Corner_plot_iband}
\end{figure}

\frig{As for the g- and r-bands, the analysis prefers the  DRW, Modified DRW, and QPO models to the periodic one, with a slight preference for the DRW.
The posterior distribution of the period found in the Periodic model becomes bimodal (see Fig.~\ref{fig:Corner_plot_iband}), explaining the large uncertainties in the predictive posterior (see Fig.~\ref{fig:best_fit_iband} panel d), with the smallest (and highest likelihood) period ($\simeq448$ days),  being closer, but still larger, to that reported by \cite[][]{Fatovic2025} than the corresponding highest peaks found for the g- and r-bands.}

\frig{As commented in Appendix~\ref{App:analysis_on_r_band}, the lower evidence for the periodic model and, within the periodic model, the lack of a consistent period in adjacent bands and the bimodal distribution of the periods further strengthen our conclusion about the lack of periodicity in the current data of J2320+0024.}

\begin{table*}
	\centering

 \caption{Summary of the different model parameters for the fit to the i-band light curve of SDSS J2320+0024. }
	\label{tab:parameters_iband}
	{\begin{tabular}{llcccccc} 
        \hline\vspace{-0.75em}\\
		Description & Name & Prior range & DRW & Modified DRW & QPO & Periodic \vspace{0.3em}\\
        \hline\vspace{-1em}\\
        \multicolumn{7}{c}{\tiny \bf 50th, 16th, 84th Percentiles}\\ \hline\vspace{-0.5em}\\
		Scale length & $\log_{10}{\tau}$ & $[-3,1.7]$ & $-1.2^{+0.3}_{-0.3}$ & $-1.2^{+0.3}_{-0.4}$ & $-1.2^{+0.4}_{-0.3}$ & -\vspace{0.2em}\\
		Dispersion & $\log_{10}{\sigma}$ & $[-3,1.7]$ & $0.6^{+0.2}_{-0.1}$ & $0.6^{+0.2}_{-0.2}$ & $0.6^{+0.2}_{-0.1}$ & $0.1^{+0.3}_{-0.2}$ \vspace{0.2em}\\
		Slope & $\gamma$ & $[0.5,5]$ & - & $1.2^{+0.7}_{-0.5}$ & - & - \vspace{0.2em}\\
		Period & $\log_{10}{T}$ & $[-2,2]$ & - & - & $0.6^{+1}_{-0.9}$ & $-0.15^{+0.01}_{-0.6}$\vspace{0.3em}\\
        \hline\vspace{-1em}\\
        \multicolumn{7}{c}{\tiny \bf Maximum Likelihood}\\ \hline\vspace{-0.5em}\\
		Scale length & $\log_{10}{\tau}$ & $[-3,1.7]$ & $-1.19$ & $-1.16$ & $-1.12$ & -\vspace{0.2em}\\
		Dispersion & $\log_{10}{\sigma}$ & $[-3,1.7]$ & $0.63$ & $0.90$ & $0.58$ & $-0.16$ \vspace{0.2em}\\
		Slope & $\gamma$ & $[0.5,5]$ & - & $2.1$ & - & - \vspace{0.2em}\\
		Period & $\log_{10}{T}$ & $[-2,2]$ & - & - & $-0.21$ & $-0.76$ \vspace{0.2em}\\
        
        \hline\vspace{-0.75em}\\
            log evidence & $\log{Z}$ & $-$ & -73.0 & -74.1 & -73.5 & -124.3 &\vspace{0.2em}\\
        \hline
        
	\end{tabular}
    \tablefoot{From left to right, the columns provide a brief description of each parameter, its reference name as used in this work, the assumed prior range, and the best-fit values with their credibility intervals for all the different models. The first table block reports the parameters estimated as the median, 16th, and 84th percentiles of the posterior distribution, while the second table block reports the set of parameters that maximise the likelihood. All the priors are log-uniform, and all the parameters are in dimensionless units since the data have been standardised before fitting. The last row reports the Bayesian log evidence of each model clearly disfavoring the Periodic model.}}

\end{table*}

\section{Quality of the fit}
\label{app:histograms}
The analysis presented in this paper is based on the Bayes ratio that, as already discussed in the paper, is the most robust approach to model selection and model comparison. However, this statement is true only under the assumption that all the models being compared are a reasonably good description of the data. To check this, we report the standardised residuals (i.e. (model-data)/error) for the same models presented in Fig.~\ref{fig:best_fit}, testing whether they are consistent with a normal distribution of zero mean and unit dispersion. For each model, we provide the residuals using either the median model (blue) or the model maximizing the likelihood (red), also reporting the mean ($\mu$) and the standard deviation ($\sigma$) of the residuals.  

In the case of good modelling of the data, the standardised residuals are expected to follow the normal distribution with $\mu=0$ and $\sigma=1$. To quantitatively asses this we performed a Kolmogorov-Smirnov test \citep[][]{Smirnov} founding that the DRW and the Modified DRW residuals always follow the expected distribution (p-value larger than the commonly adopted threshold of 0.05), the Periodic model residuals are never normally distributed while the hypothesis of a normal distribution for the QPO model can be accepted only in the case of the maximum likelihood model being compared with the data. Notably, even in this case, the p-value of the QPO model is smaller than that of the DRW or Modified DRW model.

\begin{figure}
    \centering
    \includegraphics[width=0.9\linewidth,trim={0 0 0 1.25cm}]{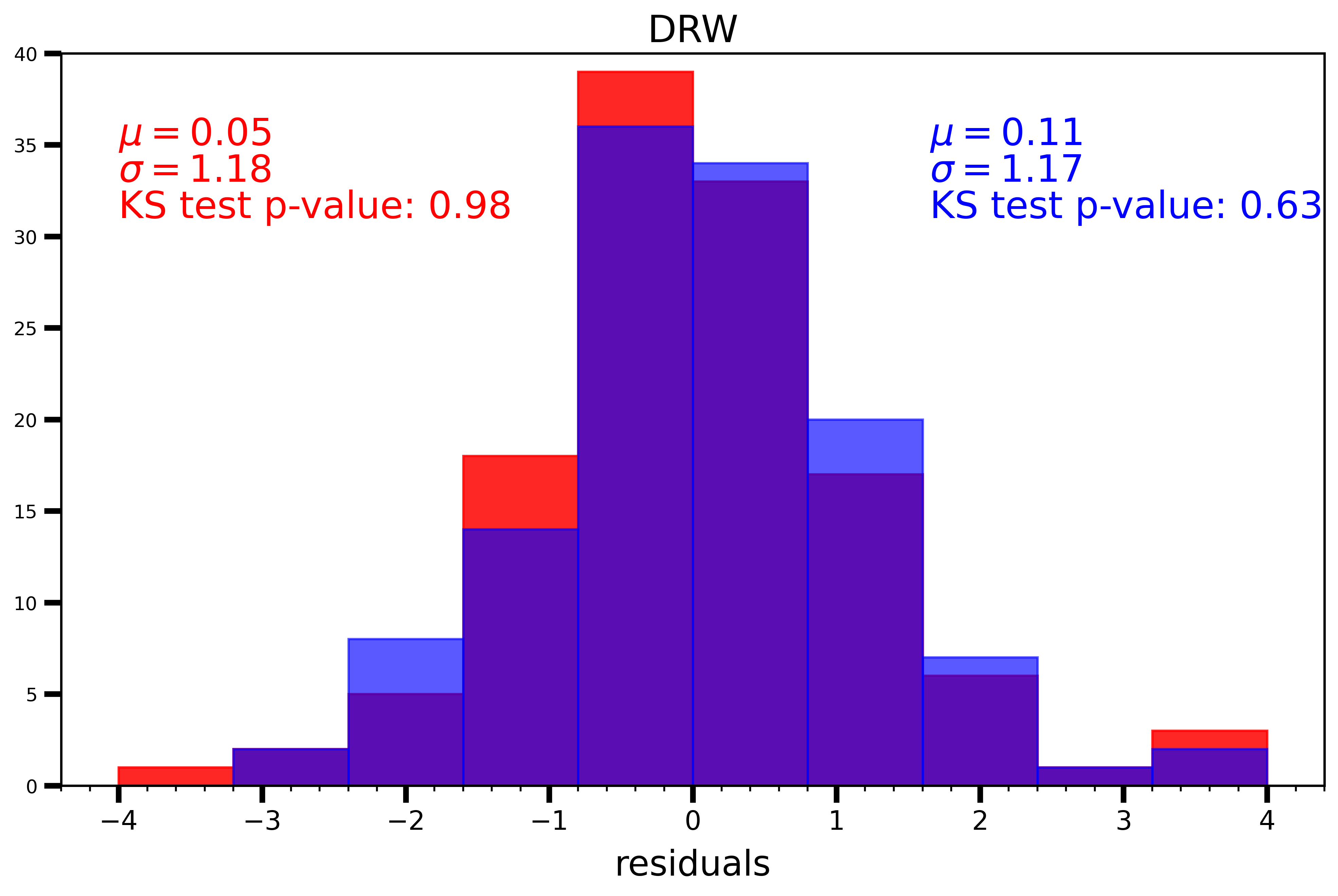}
    \includegraphics[width=0.9\linewidth]{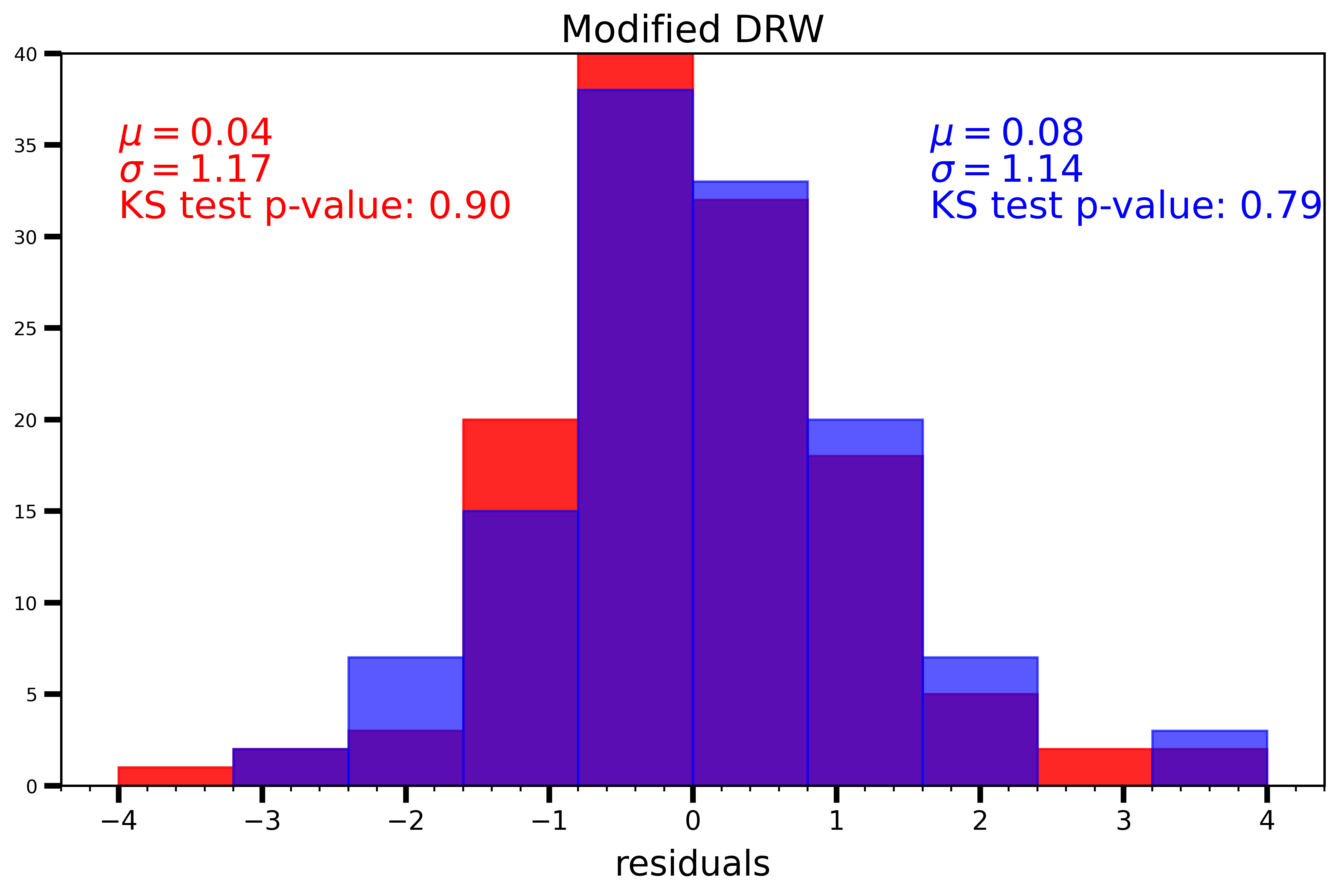}
    \includegraphics[width=0.9\linewidth]{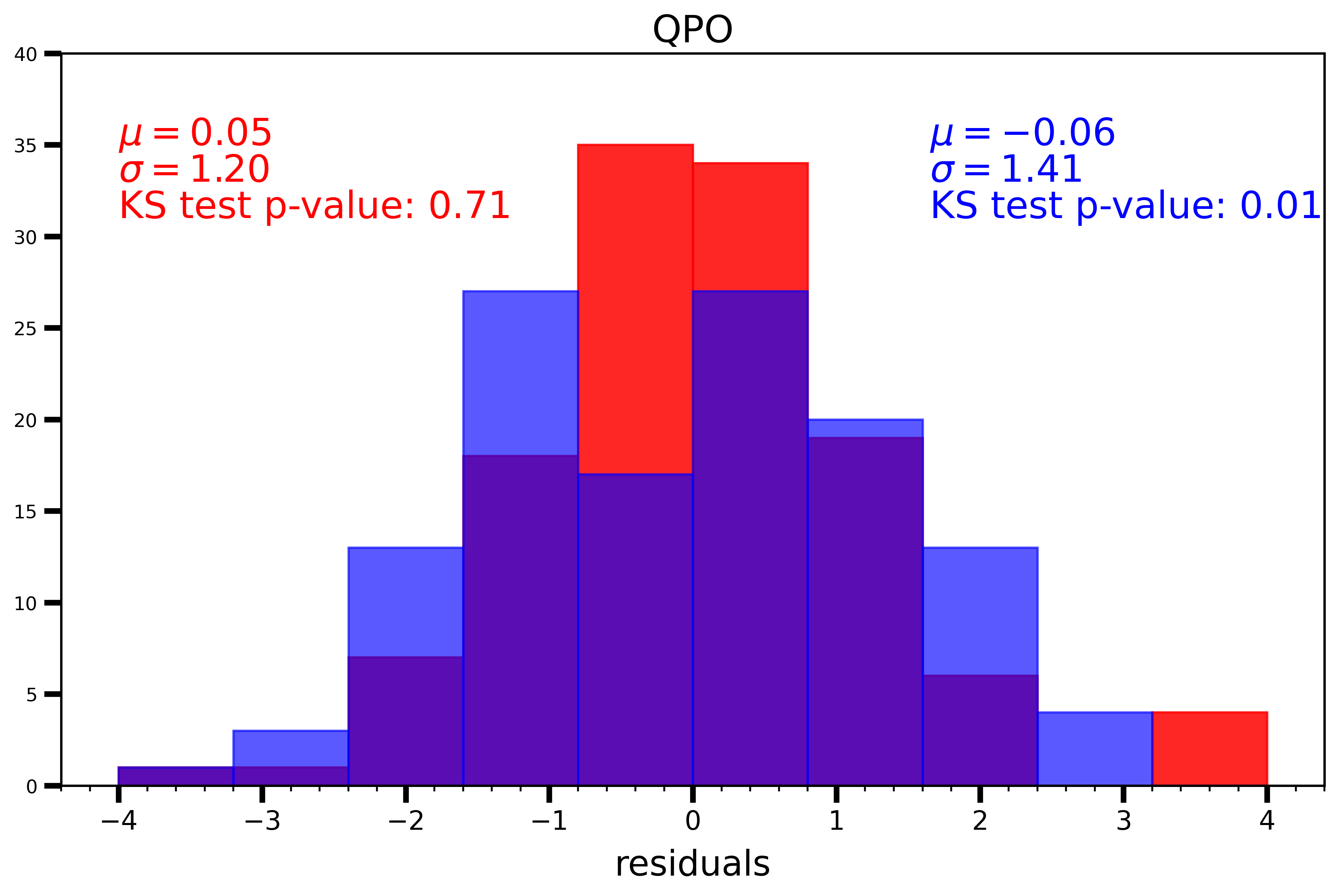}
    \includegraphics[width=0.9\linewidth,trim={0 0.8cm 0 0}]{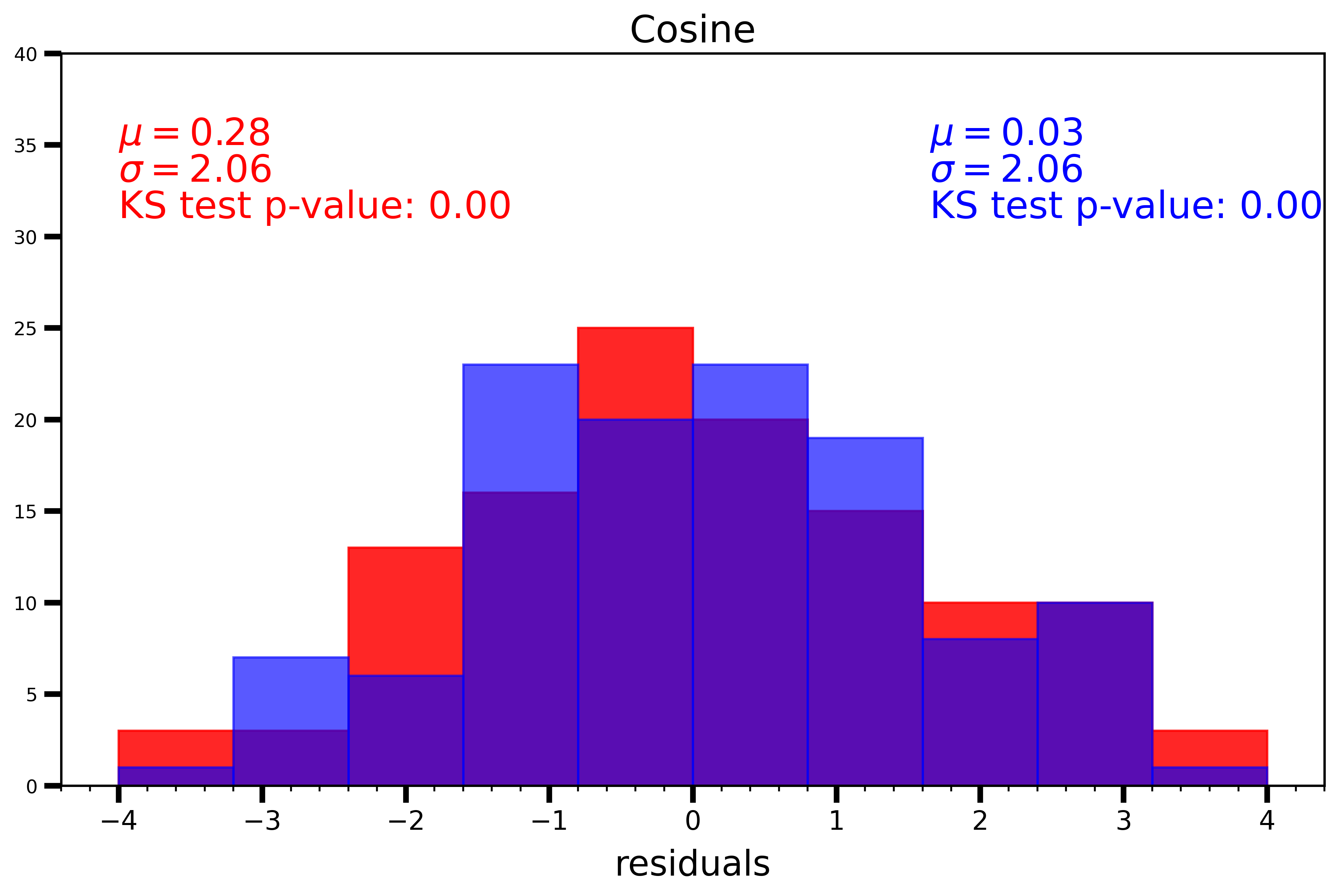}
    \caption{Histograms of the residuals (model-data)/error for the models shown in Fig.~\ref{fig:best_fit}. Each plot also reports the mean, standard deviation, and the p-values of a Kolmogorov–Smirnov test. The color coding follows the one reported in Fig.~\ref{fig:best_fit}: blue for the residual of the median model, while red for the model which maximises the likelihood. From top to bottom, we show the DRW, the modified DRW, the QPO, and the Periodic models.}
    \label{fig:best_fit_histograms}
\end{figure}

\end{appendix}


\label{lastpage}
\end{document}